\documentclass[11pt]{article}	

\usepackage[margin=.75in]{geometry}
\usepackage[utf8]{inputenc}
\usepackage[blocks]{authblk}
\usepackage{hyperref}

\usepackage{graphicx}
\usepackage{lmodern} 
\usepackage{microtype}
\usepackage{amsmath}
\usepackage{amssymb}
\usepackage{natbib, url}
\usepackage{blindtext} 
\usepackage{xcolor,colortbl}
\usepackage{blkarray}
\usepackage{setspace}
\usepackage{subcaption}
\usepackage{algorithm}
\usepackage{enumitem}
\usepackage{algpseudocode}

\usepackage{subcaption}
\usepackage{subcaption}
\usepackage{float}

\begin{document}


\title{Hierarchical Absorbing Markov Chain Models for Pooling Sparse Combat Sequence Data}
\author[1,*]{Brigham Halverson}
\author[2]{Matthew Heiner}
\author[3]{Leonardo Lamas}
\author[4]{Gilbert Fellingham}

\affil[1]{Department of Statistics, Texas A\&M University, College Station, USA, \textit{e-mail}: brighalvy@tamu.edu}
\affil[2]{Department of Statistics, Brigham Young University, Provo, USA, \textit{e-mail}: heiner@stat.byu.edu}
\affil[3]{LabEsporte, Faculty of Physical Education, University of Brasília, Brazil}
\affil[4]{Department of Statistics, Brigham Young University, Provo, USA, \textit{e-mail}: gwf@stat.byu.edu}

\date{} 


\maketitle

\begingroup
  \renewcommand{\thefootnote}{\fnsymbol{footnote}}
  \footnotetext[1]{Corresponding author: brighalvy@tamu.edu}
\endgroup
	
\abstract{By viewing tactics in no-gi Brazilian jiu-jitsu (NGJJ) as discrete states, we estimate the transition probabilities of tactical progression, while simultaneously exploring the effects of weight categories on these dynamics with a hierarchical absorbing Markov chain model. We further cluster the weight categories and borrow strength using the Ewens-Pitman attraction of Dahl, Day and Tsai. Random partitions allow us to explore the underlying groupings of these weight categories in terms of their transition patterns. The hierarchical structure enables us to obtain an overall mean estimate of the transition probability matrix along with the individual estimates of the weight categories. Through simulations we find that the proposed model is more flexible than hierarchical methods with no partition or fully-pooled alternatives while sacrificing very little efficiency under model misspecification. Applying this method to NGJJ, we found a back-take is the tactic most likely to lead to achieving submission across all weight categories. In addition, we discovered the lightest (66kg) and heaviest (99kg+) weight categories vary significantly from the others in long-term tactical progression.}


\section{Introduction}


This study aims to examine group specific tactical progression in no-gi Brazilian jiu-jitsu (NGJJ), providing statistical inference for the sport's dynamics through hierarchical modeling. NGJJ is a growing combat sport which shares many similarities with mixed martial arts (MMA). However, unlike MMA, NGJJ primarily focuses on grappling and submission-based strategies rather than striking. In a NGJJ match, athletes strive to score points by performing certain tactics successfully, or to force their opponent into submission and attain victory \citep{lamas2023no}. The progression of tactics within a match can be viewed as a sequence of discrete events (i.e.: $\text{PGD}1 \rightarrow \text{TKDA}1 \rightarrow \text{TKD}2 \rightarrow \text{BTK}1 \rightarrow \text{SUB}2$; tactic codes are described in Table \ref{tab:tbl1}). Within NGJJ there are weight categories of competition to make sure one athlete doesn't have a weight advantage over another. We aim to account for the variation in tactical progression introduced by these weight categories.  We model tactical transitions using an absorbing Markov chain to estimate the category-specific transition probabilities, such as the probability of transitioning from a back-take to a take-down in the 66kg weight category. Modeling the category-specific transition matrix becomes challenging due to sparsity within the data. We overcome this with the introduction of a random partition into the model.

A Markov chain is a probabilistic model used to analyze sequences of discrete events. The transition from one state to another depends only on the current state. The transition probability matrix ($\boldsymbol{P}$) provides insights into how frequently one state will lead to another. An absorbing Markov chain extends this concept by introducing absorbing states. Once an absorbing state is achieved, the state never changes, i.e. the sequence terminates \citep{lawler2006introduction}. In NGJJ we define our absorbing states as reset (RST), defined as an interruption of the combat sequence, or a submission (SUB) by either opponent. These outcomes either end the match or end the current sequence of tactics.

Data was collected from male athletes of the Abu Dhabi Combat Club (ADCC) – World Submission Fighting Championship in 2019. The raw data contained 93 knockout matches, over 90 athletes across weight categories: 66kg (16 athletes), 77kg (16 athletes), 88kg (16 athletes), 99kg (14 athletes), and +99kg (16 athletes) \citep{lamas2023no}. Within the data there is an additional weight category (absolute) with no weight limit, this was removed from our analysis to keep the sequential nature of the other categories and to avoid confounding.  The final data for analysis consists of 244 sequences of successive tactical moves across the five weight categories of competition. (Tactical moves are annotated in Table~\ref{tab:tbl1}) We detail the breakdown of tactical sequences in Table~\ref{tab:data_description}. This table shows how 66kg contains more sequences than the rest despite having a similar number of athletes and matches as the rest. Also notice that variation in sequence length goes down as weight goes up hinting at the presence of tactical differences between weight categories. We aim to explore how these tactical transitions in $\boldsymbol{P}$ change with weight category. We do so by incorporating a hierarchical structure into our model. 

Our NGJJ data are sparse, meaning that some tactical transitions appear frequently in some weight categories but are entirely absent in others. This sparsity eliminates the possibility of evaluating each weight category separately. To address the sparsity issue, we introduce hierarchical priors that allow weight categories to borrow strength from one another in estimation. To add flexibility in how strength is borrowed, we add a random partition, enabling weight categories to be grouped together based on similar tactical progressions. This random partition not only overcomes sparsity but gives us added insight into which categories have similar dynamics, furthering our ability to learn about the sport.

\subsection{Applications of absorbing Markov chains}

Absorbing Markov chains have been widely applied in sports analytics, including in combat sports. \cite{holmes2022markov} modeled the progression of MMA matches using Markov chains, estimating skill importance based on athlete and opponent characteristics \citep{holmes2022markov}. Similarly \cite{miarka2015development} applied multi-state Markov models to Judo, successfully estimating transition probabilities and large time behavior \citep{miarka2015development}. 

For NGJJ, \cite{lamas2023no} probabilistically described tactical progressions using Markov chains, estimating transition probabilities by pooling data across all weight categories \citep{lamas2023no}. They also outlined the tactical mapping that we use, as described in Table \ref{tab:tbl1}. We extend their analysis on an expanded dataset by accounting for the variation induced by the weight categories of competition, providing us with more category-specific insight into the sport.

Beyond combat sports, Markov chains 
have been applied to soccer for modeling goal \citep{woodfield2018quantifying} and successful outcome probabilities \citep{heiner2014skill}, to volleyball for evaluating skill importance \citep{florence2008skill}, to baseball for estimating batting transitions \citep{sokol2004intuitive}, to basketball \citep{lamas2015modeling}, and to tennis \citep{newton2009monte}. These applications highlight the broad use of Markovian modeling in sports.

\begin{table}[t]
\begin{center}

    \begin{tabular}{p{5cm} p{10cm}}
    \hline
        \textbf{Tactic} & \textbf{Description}
        \\
        Clinch dispute (CLI) & Both competitors are standing and seeking to move the opponent to the floor while maintaining a dominant position \\
        Pull guard (PGD) & The competitor sits down or lays down while intentionally pulling the opponent toward himself to exert leg control over the opponent (i.e., on guard position) \\
        Sweep (SWP) & On the floor, the competitor closer to the ground switches positions with the opponent and immediately seeks to exit the opponent's leg control \\
        Sweep attempt (SWPA) & The competitor attempts a sweep \\
        Take-down (TKD) & The competitor projects the opponent, keeping the opponent's back or shoulders on the floor \\
        Take-down attempt (TKDA) & The competitor attempts a take-down \\
        Guard pass (GPS) & The competitor escapes from the opponent's guard and achieves a ``lateral'' (transversal) or ``north-south'' (longitudinal controlling position \\
        Guard pass (GPSA) & The competitor attemps a guard pass \\
        Back-take (BTK) & The competitor escapes from the guard or using the guard achieving a dominant position from behind the opponent \\
        Back-take attempt (BTKA) & The competitor attempts a back-take \\
        Submission attempt (SUBA) & The competitor attempts to win the combat by performing a torsion or a choke \\
        Submission (SUB) & Combat stops when the competitor performs a successful torsion or choke on his opponent and his opponent gives up \\
    \hline
    \end{tabular}
    \caption{Definitions and notation for NGJJ combats. Adapted from Table 1 in \cite{lamas2023no}}
    \label{tab:tbl1}
\end{center}
\end{table}

\subsection{Hierarchical Multinomial-Dirichlet Models}

Extensive research has been done with Markov processes using Dirichlet priors. There have been a few hierarchical extensions to Markov chain models. \cite{fruhwirth2010model} performed model-based clustering of Markov chains. Their major difference from our work is the groups (categories) are treated as unknown. The likelihood for homogeneous Markov processes is very similar to the multinomial likelihood. \cite{nandram1998bayesian} proposed a hierarchical multinomial-Dirichlet model to classify text. Their model is outlined as follows: 
\begin{align*}
    \boldsymbol{x}_k\mid\boldsymbol{\theta}_k \sim \text{Multinomial}&\left(\boldsymbol{\theta}_k\right) \text{ for } k \in \{1,2,3,..., K\}, \\
    \boldsymbol{\theta}_k\mid \gamma, \boldsymbol{\alpha} &\sim \text{Dirichlet}\left(\gamma\boldsymbol{\alpha}\right), \\
    \boldsymbol{\alpha} \sim \text{Dirichlet}&, \text{   } \gamma \sim \text{Gamma},
\end{align*}
\noindent where $\boldsymbol{x}_k \in \mathbb{N}_0^J, \boldsymbol{\theta}_k \in \Delta_{J - 1}$, the $J - 1$ dimensional simplex, and $k$ indexes groups. The model introduces two hyperparamters $\gamma > 0$ and $\boldsymbol{\alpha} \in \Delta_{J - 1}$. The strength of pooling shared information across groups is represented through $\gamma$.  Being a compositional vector, $\boldsymbol{\alpha}$ can be interpreted as an overall mean for $\boldsymbol{\theta}$.

We will fit a similar model for each row of the Markov transition matrix $\boldsymbol{P}$ while adding an extension for random partitioning. Unlike previous approaches, this extension will be able to handle sparse data by combining data within a partition into one overall group. Our methods enable robust estimation of transition probabilities while preserving the hierarchical structure.

The remainder of this paper is organized as follows. Section 2 describes the data and methodology employed in this study. In Section 3, we present a simulation study that demonstrates the effectiveness of the proposed model. Section 4 details the application of the model to the NGJJ dataset, followed by a discussion of the results in Section 5.

\section{Methodology}

\subsection{Data}


\begin{table}[H]
\centering
\begin{tabular}{lrr}
  \hline
 Category & Num. Seq. & Avg. Seq. Length (SD) \\ 
  \hline
66kg & 81 & 9.864 (7.707) \\ 
   77kg & 65 & 8.815 (6.896) \\ 
   88kg & 55 & 7.673 (4.996) \\ 
   99kg & 23 & 6.478 (4.155) \\ 
   99kg+ & 20 & 7.950 (3.940) \\ 
   \hline
   \textbf{Total} & 244 & 8.615 (6.450) \\ 
   \hline
\end{tabular}
\caption{Brief description of the data, including the number of sequences with the average length of sequences and by category data. We see slight variation in tactical dynamics as observed in the number of sequences varies greatly by category and we can see as weight goes up the variation in sequence length goes down.}
\label{tab:data_description}
\end{table}

Let $x_h$ denote a single sequence of tactics, where $h = 1, ..., H$. Our data, $\{x_h\}_{h = 1}^H$, is the collection of sequences from 78 matches across five weight categories of competition within NGJJ. In total we observed 244 complete sequences. Detailed descriptions on the number of sequences is included in Table~\ref{tab:data_description}. We have a total of 20 transitory tactics, which are tactics a sequence can freely move between without terminating. These transitory tactics consist of 10 duplicated tactics being preformed by either athlete ``1'' or ``2''. This athlete assignment is arbitrary but remains consistent throughout a given sequence. The three absorbing states are submission (SUB1/SUB2) being performed by either athlete or a reset (RST). Table 1 outlines all of the possible tactics of progression, with a clinch (CLI) being coded as a reset (RST) for the purposes of our analysis. 

To proceed with the analysis, we organize the sequences of events into 10x23 matrices of counts $\boldsymbol{N}^{(k)}$, where the 10 rows correspond to tactic performed by a competitor, the first 10 columns correspond to transitory tactics performed by the same competitor, the next 10 to transitory tactics performed by opponents, with the last three corresponding to absorbing states, and $k$ corresponds to a weight category, $k \in \{1, 2, 3, 4, 5\}$. Hence, element $\boldsymbol{N}_{i,i'}^{(k)}$ will correspond to how many times a transition from tactic $i$ to tactic $i'$ occurs within the weight category $k$. 

To verify variation between weight categories exists, we performed exploratory analysis into the similarity between each $\boldsymbol{N}^{(k)}$. We found that certain weight categories appear to be similar while some differ from one another. Overall the existence of meaningful variation across weight categories is supported. Additional exploration was done into the variation introduced by athletes. We found that a similar amount of variation exists between the athlete specific data as that introduced by weight categories. Accounting for athlete-specific variation would require more data at the athlete level and is an area of future interest.


\subsection{Background on Absorbing Markov Chains}

The remainder of this section describes our methodology for general absorbing Markov chains with unrestricted transition probability matrix $\boldsymbol{P}$. In section 4, we specialize the structure of $\boldsymbol{P}$ to correspond with the NGJJ application.

In general absorbing Markov chains, transition probabilities are captured in the transition probability matrix $\boldsymbol{P}$ where the $p_{i,j}$ element would be the probability of transitioning from state $i$ to state $j$. The number of states can be arbitrary, $J$, and $\boldsymbol{P}$ would be $J \times J$ (we have $J = 23$). The likelihood of a Markov chain is:
\begin{equation}
    \prod_{i = 1}^J \prod_{j = 1}^J p_{i,j}^{n_{i,j}}
    \label{eq:MClike}
\end{equation}
\noindent where $n_{i,j}$ is the number of times the transition $i \rightarrow j$ is observed, as recorded in the count matrix $\boldsymbol{N}$. 

The sequences of NGJJ end when a RST or a SUB by either athlete occurs. These are our absorbing states, or terminal states. If $y$ is an absorbing state then $p_{y,x} = 0$ for $x \neq y$, $x \in \{1, 2, ..., J\}$, and $p_{y,y} = 1$.

Given a current state, $i$, the probability of ending up in state $j$ after $t$ steps is represented in the $i,j$ element of $\boldsymbol{P}^t$. Large time behavior is defined as $$\lim_{t \rightarrow \infty} \boldsymbol{P}^{t}.$$ This limit gives the probabilities of eventually reaching an absorbing state from a given initial state, and is available analytically under fairly general conditions \citep{lawler2006introduction}. To do so we decompose $\boldsymbol{P}$ as follows:
\begin{equation}
    \boldsymbol{P} = \left[
                \begin{array}{c|c}
                    \boldsymbol{Q}_{I\times I} & \boldsymbol{S}_{I \times m} \\  \hline
                    \textbf{0}_{m\times I} & \boldsymbol{I}_m
                    \end{array}
            \right]
\end{equation}
\noindent where $\boldsymbol{Q}$ represents the transitions among transient states, $\boldsymbol{S}$ represents transitions from transient states to absorbing states, and $\boldsymbol{I}_m$ is the identity matrix for absorbing states. Also note $m$ is the number of absorbing states and $I = J - m$ represents the number of transient states. The absorbing probability matrix is given by:
\begin{equation}
    \lim_{t \rightarrow \infty} \boldsymbol{P}^{t} = \left[
                \begin{array}{c|c}
                    \textbf{0} & \boldsymbol{A} \\  \hline
                    \textbf{0} & \boldsymbol{I}_m
                    \end{array} 
                \right],
\end{equation}
\noindent where $\boldsymbol{A} = (\boldsymbol{I}_I - \boldsymbol{Q})^{-1} \boldsymbol{S}$ \citep{lawler2006introduction}. In NGJJ this provides insight into the probability of an athlete ultimately reaching a reset or submission from any given tactic.

\subsection{Methods Considered}

The objective of this analysis is to develop a model for $\boldsymbol{P}^{(k)}$ for $k = 1, ..., K$ while accounting for variation introduced by $K$ weight categories. We will denote the weight category for sequence $h$ as $k_h \in \{1, 2, ..., K\}$, with $K = 5$. The likelihood for our data is expressed as follows:
\begin{equation}
    x_h \mid \boldsymbol{P}^{(k_h)}, k_h \overset{ind}{\sim} \text{Absorbing Markov Chain}(\boldsymbol{P}^{(k_h)}) \text{ for } h = 1, ..., H,
    \label{eq:orig_mc}
\end{equation}
with $k_h \in \{1, 2, \hdots, K\}$. The matrices $\{\boldsymbol{N}^{(k)}\}_{k = 1}^K$ collectively provide sufficient statistics for estimating $\{\boldsymbol{P}^{(k)}\}_{k = 1}^K$. 

We build upon previously applied methods to arrive at our final hierarchical model with group-level clustering. This section outlines those previously used methods and builds upon them to arrive at our methodology. Section 2.3.1, uses complete pooling of information across all groups; this is the widely used application of Markov chains and is the method of \cite{lamas2023no}. In Section 2.3.2, we introduce a hierarchical structure that allows for group-specific inference while still borrowing strength across groups. Finally, Section 2.3.3 extends this approach by incorporating a random partition, allowing for flexible clustering of groups while maintaining the hierarchical structure.

\subsubsection{Absorbing Markov Chain - Dirichlet Model (AMC)}

A straightforward approach to modeling $\{\boldsymbol{P}^{(k)}\}_{k = 1}^K$ follows the method outlined by \cite{lamas2023no}. In this approach, $\boldsymbol{P}^{(k)} = \boldsymbol{P}$ for all $k = 1, ..., K$ and all the data are pooled together, i.e. $\boldsymbol{n}_i = \sum_{k = 1}^K \boldsymbol{n}_i^{(k)}$, where $\boldsymbol{n}_i$ is the $i^{th}$ row of $\boldsymbol{N}$. Pooling the data as such provides a universal estimate of $\boldsymbol{P}$ regardless of group.

Each row of $\boldsymbol{P}$, $\boldsymbol{p}_i$, is modeled independently with a Dirichlet prior:
\begin{equation}
    \boldsymbol{p}_i \overset{\mathrm{ind}}{\sim} \operatorname{Dirichlet}(\boldsymbol{\xi}_i), \quad i = 1, \dots, I
\end{equation}
\noindent where $\boldsymbol{\xi}_i$ is a fixed prior shape vector. This model is fully conjugate, allowing posterior inference for each $\boldsymbol{p}_i$ to be directly obtained from a Dirichlet distribution with shape vector $\boldsymbol{n}_i + \boldsymbol{\xi}_i$.

The AMC model has low computational cost and minimizes sparsity in $\boldsymbol{N}$ by pooling all groups together. However, this approach does not account for differences among groups, which limits its applicability when group-level inferences are of interest.

\subsubsection{Hierarchical Absorbing Markov Chain Model (HAMC)}

To address the limitation of the AMC model and allow for group-specific inference, we introduce a hierarchical structure. We model the group-specific sequences as before with each row of $\boldsymbol{P}^{(k)}$, $\boldsymbol{p}_i^{(k)}$, treated independently across $i$, similar to the AMC model. However, now the Dirichlet priors on each $\boldsymbol{p}_i^{(k)}$ are connected hierarchically across $k$,
\begin{equation}
    \boldsymbol{p}_i^{(k)} \mid \boldsymbol{\alpha}_i, \gamma_i \overset{ind}{\sim} \text{Dirichlet}(\boldsymbol{\alpha}_i \gamma_i).
\end{equation}
\noindent The hyperparameters $\gamma_i$ and $\boldsymbol{\alpha}_i$ control the degree of information sharing across groups. The hyper-prior distributions are are:
\begin{align}
    \boldsymbol{\alpha}_i &\overset{ind}{\sim} \text{Dirichlet}(\frac{1}{J} \cdot \boldsymbol{1}_i) \\
    \gamma_i &\overset{iid}{\sim} \text{Gamma}(a_\gamma, b_\gamma),
\end{align}
\noindent where $b_\gamma$ is the rate. The prior on $\boldsymbol{\alpha}_i$ enforces the constraint, $\sum_{j = 1}^J \alpha_{i,j} = 1$ necessary for identifiability and allows $\boldsymbol{\alpha}_i$ to be interpreted as a probability vector. Thus $\boldsymbol{\alpha}_i$ will represent the mean transition probability vector common among all groups.  This setup lets $\gamma_i$ represent the relative strength of contribution of the pooled information from groups for row $i$.

The conjugacy of the Markov chain likelihood to the independent Dirichlet priors allows us to integrate out all the transition probabilities and model each row of $N^{(k)}$ as outlined by \cite{nandram1998bayesian}. We derive collapsed full conditional distributions of both $\boldsymbol{\alpha}_i$ and $\gamma_i$ which are free of the transition probabilities contained in $\boldsymbol{P}^{(k)}$. These derivations are outlined for the final EPA-HAMC model in Section~\ref{sec:marg_cond} of the Appendix.

With this model we are able to borrow strength across groups, perform inference on the differences in $\boldsymbol{P}^{(k)}$ among the groups, and obtain a global estimate for $\boldsymbol{P}$. Despite these advantages, sparsity of the NGJJ data presents an issue when splitting $\boldsymbol{N}$ into each $\boldsymbol{N}^{(k)}$. 

\subsubsection{Group Level Clustering (EPA-HAMC)}

To create a middle ground between the AMC and HAMC models, we allow groups with similar transition dynamics to cluster together. This is distinct from clustering each sequence, $x_h$ as their group labels $k_h$ are known. Here, we are clustering the groups (i.e. weight categories) to compare their characteristics, and to borrow strength and partially pool data across groups. This is achieved using a random partition model, where groups, $\{1, 2, ..., K\}$, are grouped into mutually exclusive clusters, $S_g \subseteq \{1, 2, ..., K\}$ with $g = 1, \hdots G$ where $1 \leq G \leq K$ such that $\cup_{g = 1}^G S_g = \{1, \hdots, K\}$. The clusters are organized into a partition, $\pi_i = \{S_{1i}, ..., S_{G,i}\}$, where all groups in a cluster share the same transition probabilities from tactic (row) $i$. We will use the same sequential sampling model as outlined in Equation \ref{eq:orig_mc}. The model is still specified independently on each row within a group, $\boldsymbol{p}_i^{(k)}$, where now we define
\begin{equation}
    \boldsymbol{p}_i^{(k)} = \sum_{g = 1}^{G_i} \boldsymbol{\theta}_i^{(g)}\cdot \boldsymbol{I}\{k \in S_{g,i}\}.
\end{equation}
The prior distributions remain similar to the HAMC model:
\begin{equation}
    \begin{split}
        \boldsymbol{\theta}_i^{(g)} \mid \boldsymbol{\alpha}_i, \gamma_i &\sim \text{Dirichlet}(\boldsymbol{\alpha}_i \gamma_i), \\
        \boldsymbol{\alpha}_i &\sim \text{Dirichlet}(\frac{1}{J} \cdot \boldsymbol{1}_J), \\
        \gamma_i &\sim \text{Gamma}(\tau, \eta).
\end{split}
\label{eq:EPA_HAMC_mod}
\end{equation}
\noindent Given any partition, $\pi_i$, the model for $\boldsymbol{\theta}_i^{(g)}$ is equivalent to that of the HAMC model. We model $\pi_i$ using a Ewens-Pitman attraction (EPA) process introduced by \cite{dahl2017random}. We write
\begin{equation}
    \pi_i \mid \beta_i, \delta_i, \boldsymbol{\sigma}_i \overset{ind}{\sim} \text{EPA}(\beta_i, \delta_i, \lambda_i, \boldsymbol{\sigma}_i),
    \label{eq:EPA_prior}
\end{equation}
\noindent with four parameters that influence the partition $\pi_i$. One benefit of the EPA prior is its ability to incorporate prior information on the similarities between the groups through the similarity metric $\lambda_i$. This is accomplished by providing pairwise ``distances'' between groups. Let $d_{x,y}$ be a distance metric between groups $x$ and $y$. Then we define similarity $\lambda_i(x,y) = d_{x,y}^{-\tau}$ with $\tau > 0$ \citep{dahl2017random}. For our model we use $\tau = 0.5$ as suggested by \cite{dahl2017random}. However, a gamma prior can be used to gain inference and tune the similarity metric $\lambda_i$. In the prior of the EPA-HAMC we encode a distance that each weight category is one away from each adjacent weight category and two away from weight categories two above or below in weight classification, and so forth. Thus, 66kg has a prior distance of four from 99kg+.

The remaining parameters ($\beta_i$, $\delta_i$, $\boldsymbol{\sigma}_i$) allow for model clustering flexibility. The mass parameter $\beta_i$ ($\beta_i > -\delta_i$) and the discount parameter $\delta_i$ ($0 \leq \delta_i < 1$) determine the prior expected number of clusters. In the special case when $\delta_i = 0$, the EPA is equivalent to the Chinese restaurant process. \cite{dahl2017random} suggests using a mixture of a point mass at zero and a beta distribution as a prior. To simplify the MCMC algorithm, we use a mixture of two beta distributions one with high density near zero:
\begin{equation}
    \delta_i \sim w\text{Beta}(.05, 20) + (1-w)\text{Beta}(a_\delta,b_\delta).
    \label{eq:beta_prior}
\end{equation}
\noindent For the mass parameter ($\beta_i$) a gamma prior will be used. The pmf of EPA is non-exchangeable and relies on a sequential allocation of groups, $\boldsymbol{\sigma}_i$. We use a uniform prior and follow the MCMC procedure given by \cite{dahl2017random}.

\subsection{Posterior Estimation}
\label{sec:post_est}

We use MCMC to approximate the posterior distributions of all parameters. We derive conditional distributions of the parameters $\pi_i$, $\boldsymbol{\alpha}_i$ and $\gamma_i$ that are free of the probability vectors in Section~\ref{sec:marg_cond} of the Appendix. These derived conditional distributions simplify the MCMC algorithm.

Due to the multivariate nature and the correlation among elements of $\boldsymbol{\alpha}_i$ in the posterior, we use generalized elliptical slice sampling \citep{nishihara2014parallel}. This requires the utilization of an appropriate pseudo-target (proposal distribution) for $\boldsymbol{\alpha}_i$. The posterior distribution of $\boldsymbol{\alpha}_i$ isn't exactly Dirichlet. However, the hierarchical interpretation suggests an approximating Dirichlet distribution can provide the basis of a successful pseudo-target. We let $\widehat{\boldsymbol{\theta}}_{i,pool} = N_i^{-1} \sum_{k = 1}^K \boldsymbol{\boldsymbol{n}}_i^{(k)}$, where $N_i = \sum_{k = 1}^K\sum_{j= 1}^Jn_{i,j}^{(k)}$. Further, let  $\widehat{\boldsymbol{\theta}}_{i,ind}^{(g)} = (N_i^{(g)})^{-1}\sum_{k = 1}^K\boldsymbol{n}_i^{(k)}\boldsymbol{I}\{k \in S_{g,i}\}$ where $N_i^{(g)} = \sum_{k = 1}^k\sum_{j = 1}^J n_{i,j}^{(k)}\boldsymbol{I}\{k \in S_{g,i}\}$. Additionally let $\overline{\boldsymbol{\theta}}_{i,ind} = G_i^{-1}\sum_{g = 1}^{G_i}\widehat{\boldsymbol{\theta}}_{i,ind}^{(g)}$ and $\overline{N}_i = {G_i}^{-1}\sum_{g = 1}^{G_i} N_i^{(g)}$. For large values of $\gamma_i$, the similarity between groups is high and thus $\{\boldsymbol{\theta}_i^{(g)}\}$ concentrate near $\boldsymbol{\alpha}_i$ making $\widehat{\boldsymbol{\theta}}_{i,pool}$ a good estimator. Conversely small $\gamma_i$ results in unique $\{\boldsymbol{\theta}_i^{(g)}\}$, for which $\{\widehat{\boldsymbol{\theta}}_{i,ind}^{(g)}\}$ provide essentially independent observations informing $\boldsymbol{\alpha}_i$. Scaling the two into pseudo-counts yields $$\widehat{\alpha}_i^{mix} = \frac{1}{J}\boldsymbol{1}_i + \gamma_i\widehat{\boldsymbol{\theta}}_{i,pool} + \overline{N}_i \overline{\boldsymbol{\theta}}_{i,ind}.$$ 

To enable elliptical slice sampling, we use a multivariate logistic transformation $\eta_{i,j} = \log(\alpha_{i,j}/\alpha_{i,J}) \in \mathbb{R}.$ The pseudo-target for $\boldsymbol{\eta}_i = (\eta_{i,j}, \hdots, \eta_{i, J-1})^\top$ is derived from the multivariate Gaussian distribution with minimal Kullback-Leibler distance from the Dirichlet pseudo-target for $\boldsymbol{\alpha}_i$ \citep{aitchison1986}, with mean and covariance given by 
\begin{align*}
    \widehat{\mu}_{i,j} &= \psi(\widehat{\alpha}_{i,j}^{mix})- \psi(\widehat{\alpha}_{i,J}^{mix}) \\
    \widehat{\Sigma}^{(i)}_{j,j'} &= \psi'(\widehat{\alpha}_{i,J}^{mix}) + \psi'(\widehat{\alpha}_{i,j}^{mix})\boldsymbol{I}\{j = j'\},
\end{align*}
for $j, j' \in \{1, \hdots, J - 1\}$, where $\psi()$ and $\psi'()$ are the digamma and trigamma functions.

Our generalized elliptical slice sampler (GESS, \citealp{nishihara2014parallel}) uses the importance ratio $$\frac{p_{\boldsymbol{\eta}_i}(\boldsymbol{\eta} \mid \cdot)}{t_\nu(\boldsymbol{\eta}; \widehat{\boldsymbol{\mu}}_i, \widehat{\Sigma}^{(i)})},$$ where $p_{\boldsymbol{\eta}_i}(\boldsymbol{\eta} \mid \cdot)$ is the full conditional distribution of $\boldsymbol{\eta}_i$, and $t_{\nu}$ is a Student's $t$ distribution with $\nu = 9$ degrees of freedom. The heavy tails of the $t$ distribution guarantee that this importance ratio is bounded above for all $\boldsymbol{\eta} \in \mathbb{R}^{J - 1}$.

We use Metropolis-Hastings, slice sampling, or Gibbs sampling methods to draw from the full conditional distributions of the remaining parameters. A full MCMC algorithm is outlined in Section D of the Supplementary Material. While the high number of parameters and the dimensionality of the respective spaces naturally increase the per iteration run time our MCMC algorithm achieve excellent mixing and reliable posterior convergence as demonstrated in Section~\ref{sec:simulation}.

\section{Simulation Study}
\label{sec:simulation}

The purpose of the simulation study is to prove the concept and effectiveness of both the HAMC and EPA-HAMC methodology. To do so data was simulated under five \textit{scenarios} for partition structure;

\noindent \textit{Scenario 1:} All group are part of the same cluster. (AMC model is the truth)

\noindent \textit{Scenario 2:} All groups are in their own cluster. (HAMC model is the truth)

\noindent \textit{Scenario 3:} There are three distinct clusters, consistent across each row, with even distribution of groups among each partition. (EPA-HAMC model is the truth)

\noindent \textit{Scenario 4:} Each row has three distinct clusters with even distribution of groups among each partition but the cluster assignment varies across each row. (EPA-HAMC model is the truth)

\noindent \textit{Scenario 5:} There are again three clusters. However, there is an uneven distribution of groups among the clusters. The clusters are different across each row. (EPA-HAMC model is the truth)

To evaluate the results, we split the simulation into three sets. The first two are given here in detail, the third is deferred to Section E of the Supplementary Material. 
\begin{itemize}[label={}]
\item \textbf{Simulation 1} uses $I = 30$ transitory states and $K = 5$ groups under \textit{Scenarios 1, 2, \& 3} and evaluates each respective model's reconstruction of $\boldsymbol{P}^{(k)}$ through $$\text{MSE}_P = \sum_{k = 1}^K||\text{vec}({\boldsymbol{P}}^{(k)}) - \text{vec}(\widehat{{\boldsymbol{P}}}^{(k)})||_2^2/KIJ,$$ $$\text{MSE}_A = \sum_{k = 1}^K||\text{vec}({\boldsymbol{A}}^{(k)}) - \text{vec}(\widehat{{\boldsymbol{A}}}^{(k)})||_2^2/KIJ,$$ where $\boldsymbol{A} = \lim_{t \rightarrow \infty} \boldsymbol{P}^t$, and $\widehat{{\boldsymbol{P}}}^{(k)}$ and $\widehat{{\boldsymbol{A}}}^{(k)}$ are the posterior means. We also observe the coverage and interval widths of 95\% posterior credible intervals of $\boldsymbol{P}^{(k)}$. Results are given in Section~\ref{sec:sim1}.
\item \textbf{Simulation 2} demonstrates that the EPA-HAMC model can accurately recover the number of partitions among the groups. Data are simulated using $I = 30$ and $K = 5$ and simulated under each \textit{Scenario}. Results for this simulation are given in Section~\ref{sec:sim2}
\end{itemize}

Each \textbf{simulation} uses the following procedure to generate the count matrix $\boldsymbol{N}^{(k)}$. 
\begin{enumerate}
    \item First the partition for each row $i$ is specified.
    \item $\gamma_i = 10$ and $\boldsymbol{\alpha}_i$ is simulated from a Dirichlet distribution.
    \item Given the grouping for each row, $\boldsymbol{p}_i^{(k)}$ is simulated from its prior distribution.
    \item Then using the constructed $P^{(k)}$, $N^{(k)}$ is simulated starting a sequence in each transitory state 10 times. Each sequence is simulated until absorption.
\end{enumerate}
In each case the number of absorbing states is three. In each case we obtain 10,000 posterior draws after a burn-in period of 500 draws. The EPA-HAMC model is thinned to every 5 iterations to obtain the same 10,000 posterior samples. Each study is replicated 50 times. 

\subsection{Simulation 1: Evaluation of EPA-HAMC, HAMC, AMC}
\label{sec:sim1}

\begin{figure}[H]
    \centering
    \begin{subfigure}[b]{\textwidth}
        \centering
        \includegraphics[width=\textwidth]{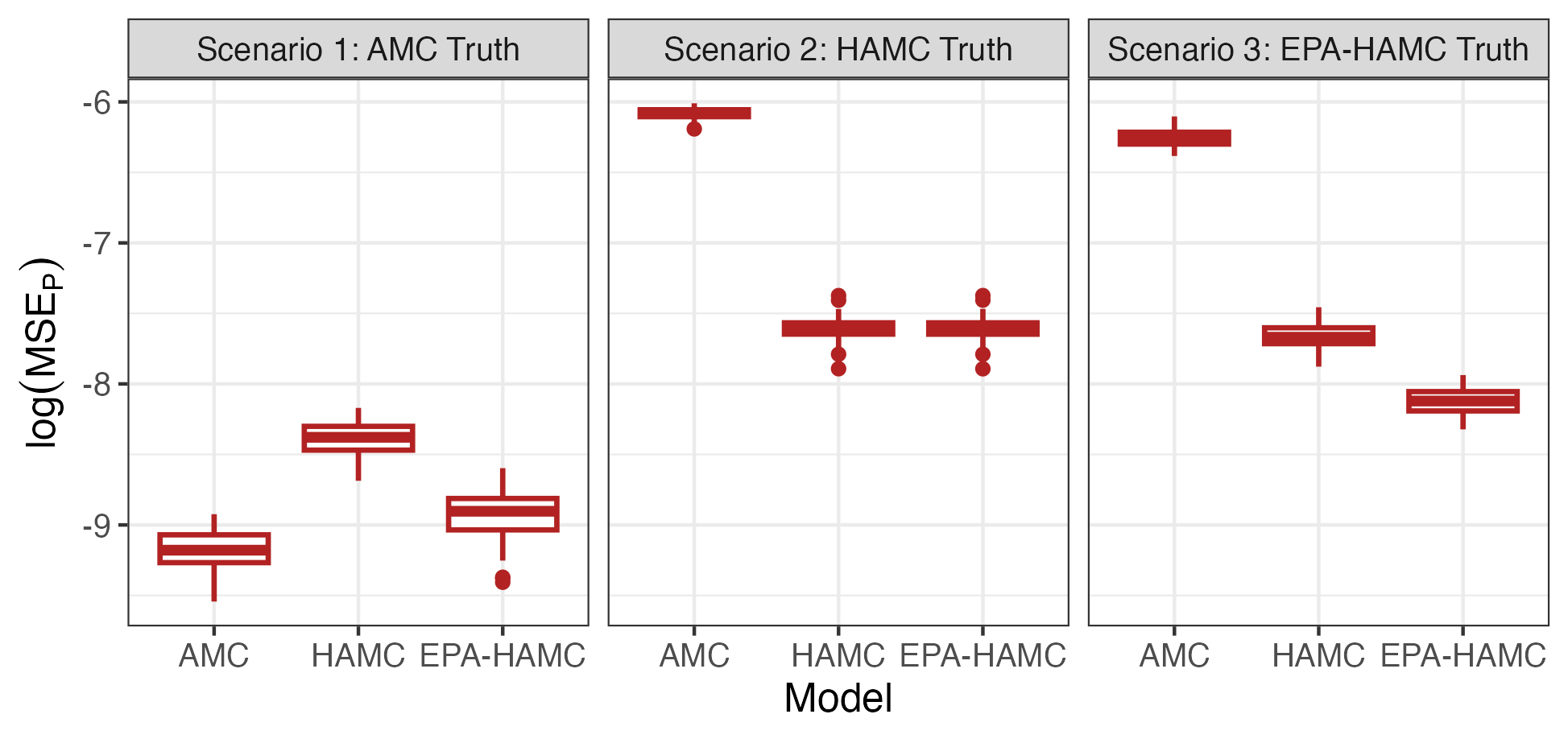}
        \caption{log(MSE$_{P}$)}
        \label{fig:MSE_P}
    \end{subfigure}
    
    \vspace{0.3cm} 
    
    \begin{subfigure}[b]{\textwidth}
        \centering
         \includegraphics[width=\textwidth]{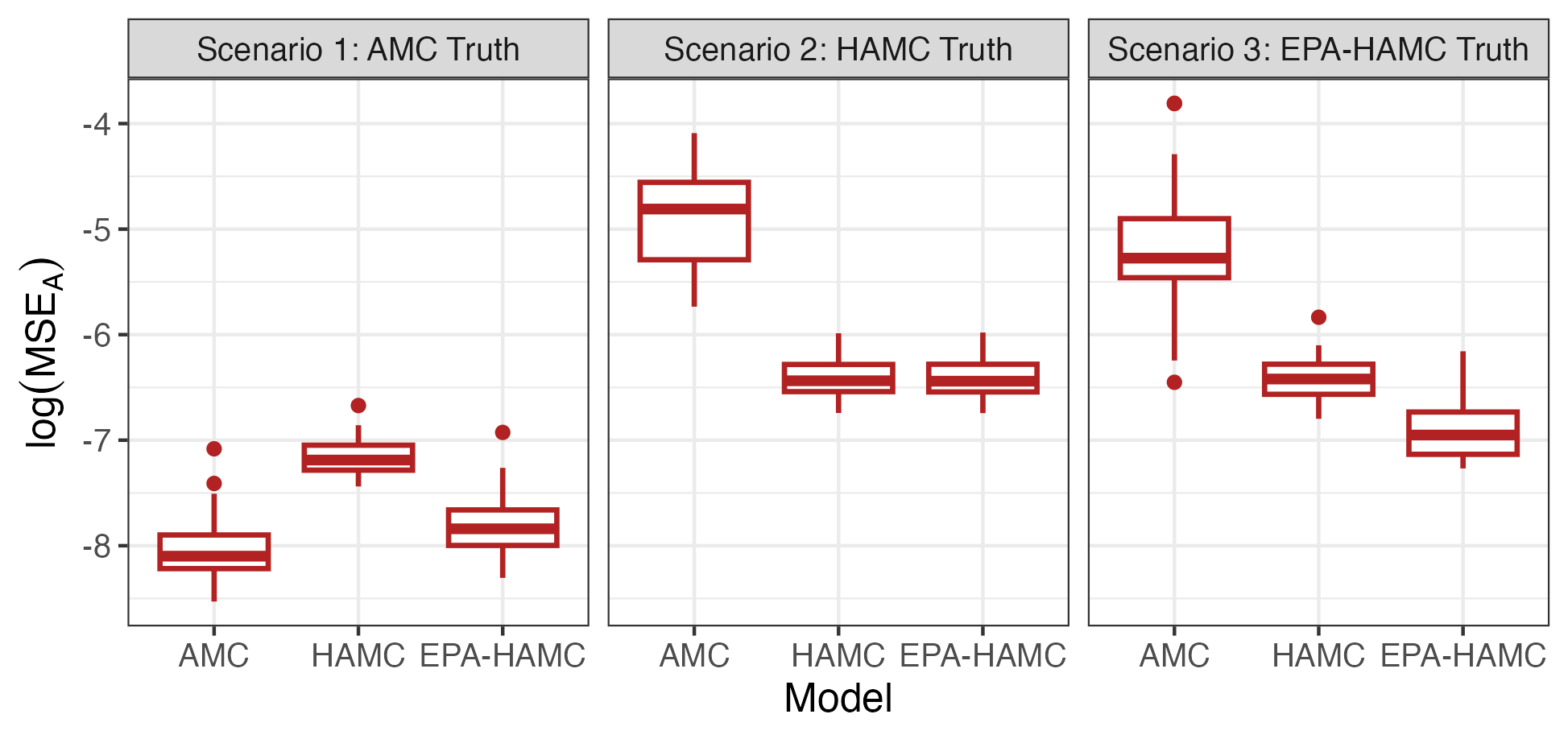}
        \caption{log(MSE$_{A}$)}
        \label{fig:MSE_A}
    \end{subfigure}

    \caption{\ref{fig:MSE_P} is boxplots of log(MSE$_P$) across the 50 replicates. While \ref{fig:MSE_A} is boxplots log(MSE$_A$). In each \textit{scenario} the true data generating model performs the best, however, we see that EPA-HAMC performs competitively in each \textit{scenario} for both immediate and long-term behavior.}
    \label{fig:MSE_sim1}
\end{figure}

Figure \ref{fig:MSE_sim1} illustrates the distribution of $\text{MSE}_P$ and $\text{MSE}_A$ across the initial three experimental scenarios. Unsurprisingly, the true data-generating model establishes the lower bound for error in both immediate ($\text{MSE}_P$) and long-term behavior ($\text{MSE}_A$). Crucially, however, the proposed EPA-HAMC framework exhibits competitive performance across all \textit{scenarios}, demonstrating particularly robust accuracy. This resilience indicates that adopting EPA-HAMC imposes low model misspecification costs, even when the data originates from one of two extreme partitioning structures. While the standard HAMC is flexible and borrows strength, the EPA-HAMC model can pool more strongly when appropriate which we observe through the comparative performance. Moreover, in settings where the true underlying model is the standard HAMC (\textit{Scenario 2}), EPA-HAMC’s long-term behavior is nearly identical to the ground truth, underscoring its capability to capture complex partitioning dynamics without overfitting.

\begin{figure}[H]
    \centering
    \includegraphics[width = \linewidth]{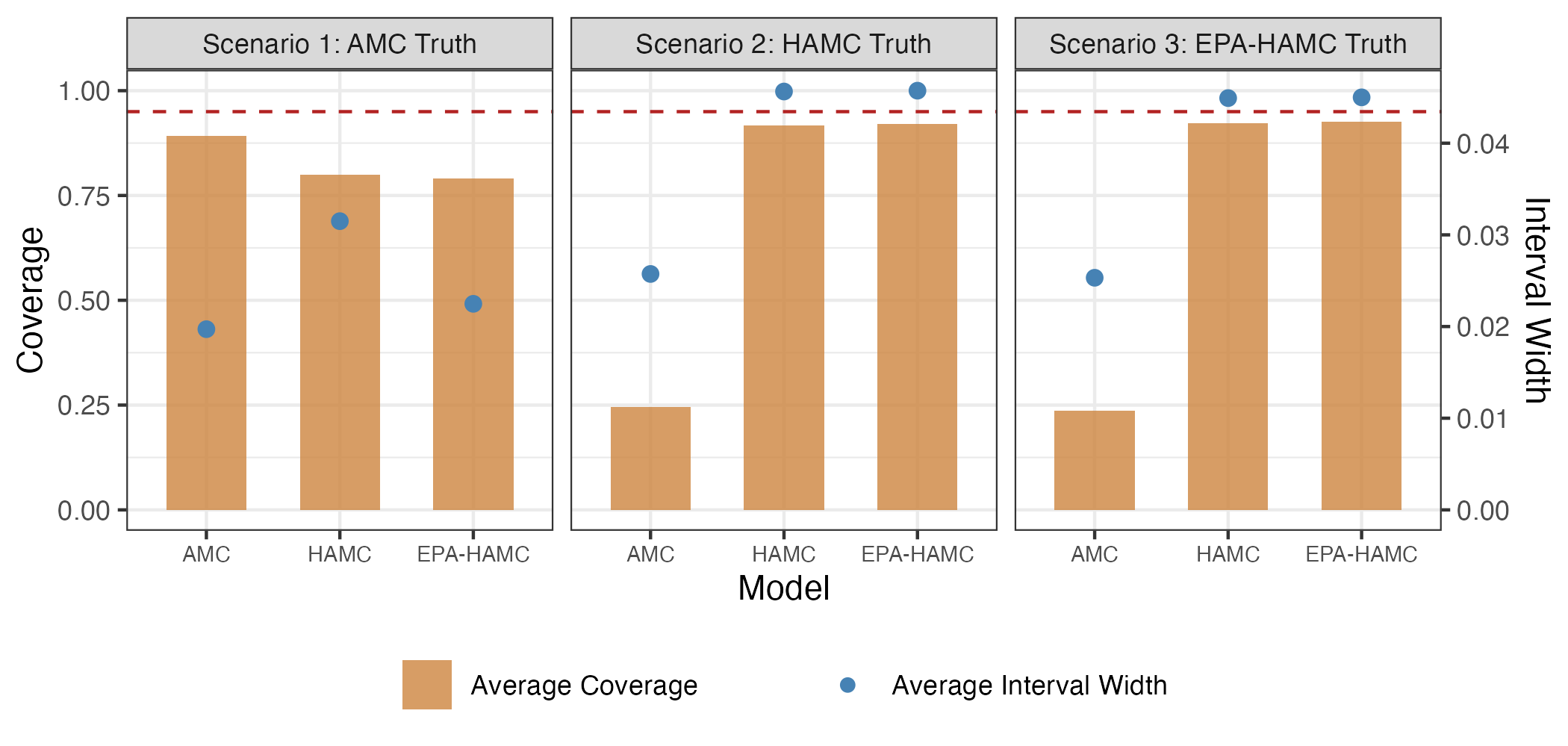}
    \caption{Average coverage and interval width of 95\% posterior credible intervals of $P^{(k)}$ for each \textit{Scenario 1, 2, \& 3} across the three methods (AMC, HAMC, EPA-HAMC). The red dotted line depicts the 95\% target coverage for the posterior intervals. HAMC and EPA-HAMC perform well on coverage under each \textit{Scenario} where the AMC has a steep drop off when there is more than one cluster. EPA-HAMC does well on interval width in each scenario compared to the true method. While AMC always has the smallest interval, it doesn't always capture the truth when partitions are present. }
    \label{fig:sim_coverage}
\end{figure}

We observe from Figure~\ref{fig:sim_coverage} that the empirical coverage and interval widths mirror the performance seen in the $\text{MSE}_P$ and $\text{MSE}_A$ metrics, where each respective method expectedly achieves optimal performance when evaluated under its data-generating \textit{scenario}. However, a deeper dive reveals critical trade-offs in uncertainty quantification; the AMC model consistently yields the narrowest interval widths. However, it suffers from a severe drop-off in coverage well below the nominal 95\% target line in any \textit{scenario} containing more than a single partition as would be expected. Conversely, the proposed EPA-HAMC method consistently performs on par with the true data-generating model across all evaluated \emph{scenarios}, successfully maintaining coverage without suffering from the overly conservative or inflated interval widths. Ultimately, these results demonstrate that the EPA-HAMC model misclassification costs are remarkably low in terms of posterior inference.

\subsection{Simulation 2: Evaluation of Partition Estimation}
\label{sec:sim2}

\begin{table}[H]
\centering
\begin{tabular}{l|rr}
\textit{Scenario} & True \# of Clusters & Mean (sd) \\ 
\hline
1 & 1 & 1.48 (0.15) \\
2 & 5 & 5.00 (0.00) \\
3 & 3 & 3.01 (0.01) \\
4 & 3 & 3.00 (0.01) \\
5 & 3 & 3.00 (0.00) \\
\end{tabular}
\caption{Posterior mean number of clusters for the EPA-HAMC model under varying levels of complexity for the true number and structure of clusters (\emph{scenario}). Each result is summarized across 50 replicates showing the robustness of accurately recovering the true number of clusters.}
\label{tab:num_results}
\end{table}

For each \textit{scenario} the EPA-HAMC model was fit and at each post-burn in iteration the number of partitions was monitored. Then, the average posterior number of partitions was collected from each replicate. Across all \textit{scenarios} the results observed in Table~\ref{tab:num_results} show we are consistently recovering the true number of clusters with a high degree of precision. These results show the underlying cluster selection mechanism remains highly identifiable, even under varying \textit{scenario} complexities.


\section{NGJJ Results}

In the NGJJ analysis, each $\boldsymbol{P}^{(k)}$ will be a 23x23 matrix of probabilities. Due to the nature of the Markov model, the rows corresponding to the absorbing states have a probability one in the $(i,i)$ element of the matrix. Thus we estimate only the first $I = 20$ rows of $\boldsymbol{P}^{(k)}$ corresponding with the trasient tactics. In NGJJ certain transitions are impossible, and we fix their probabilities as structural zeros. The number of possible transitions for row $i$ is denoted by $J_i$. These constraints are easily accommodated because we treat each row of $\boldsymbol{P}^{(k)}$ independently.

The numbers that are assigned to the athletes within a sequence are arbitrary; in other words it doesn't matter who is ``1'' or who is ``2''. Thus the transition probability of BTKA1 $\rightarrow$ BTK1, for instance, is the same as the transition probability of BTKA2 $\rightarrow$ BTK2. This likewise applies to between competitors so that, for example, BTK1 $\rightarrow$ TKDA2 is equivalent to BTK2 $\rightarrow$ TKDA1. Thus our model only needs to estimate  $I = 10$ unique rows of $\boldsymbol{P}^{(k)}$. To assist in model fitting, we combine the counts of equivalent moves within the count matrix $\boldsymbol{N}^{(k)}$. Section A of the Supplementary Material depicts an example of this symmetry and of the structure of $\boldsymbol{P}$ using six total transient states (three repeated states performed by either person ``1'' or ``2'') and two absorbing states.

\begin{figure}
    \centering
    \includegraphics[width=\linewidth]{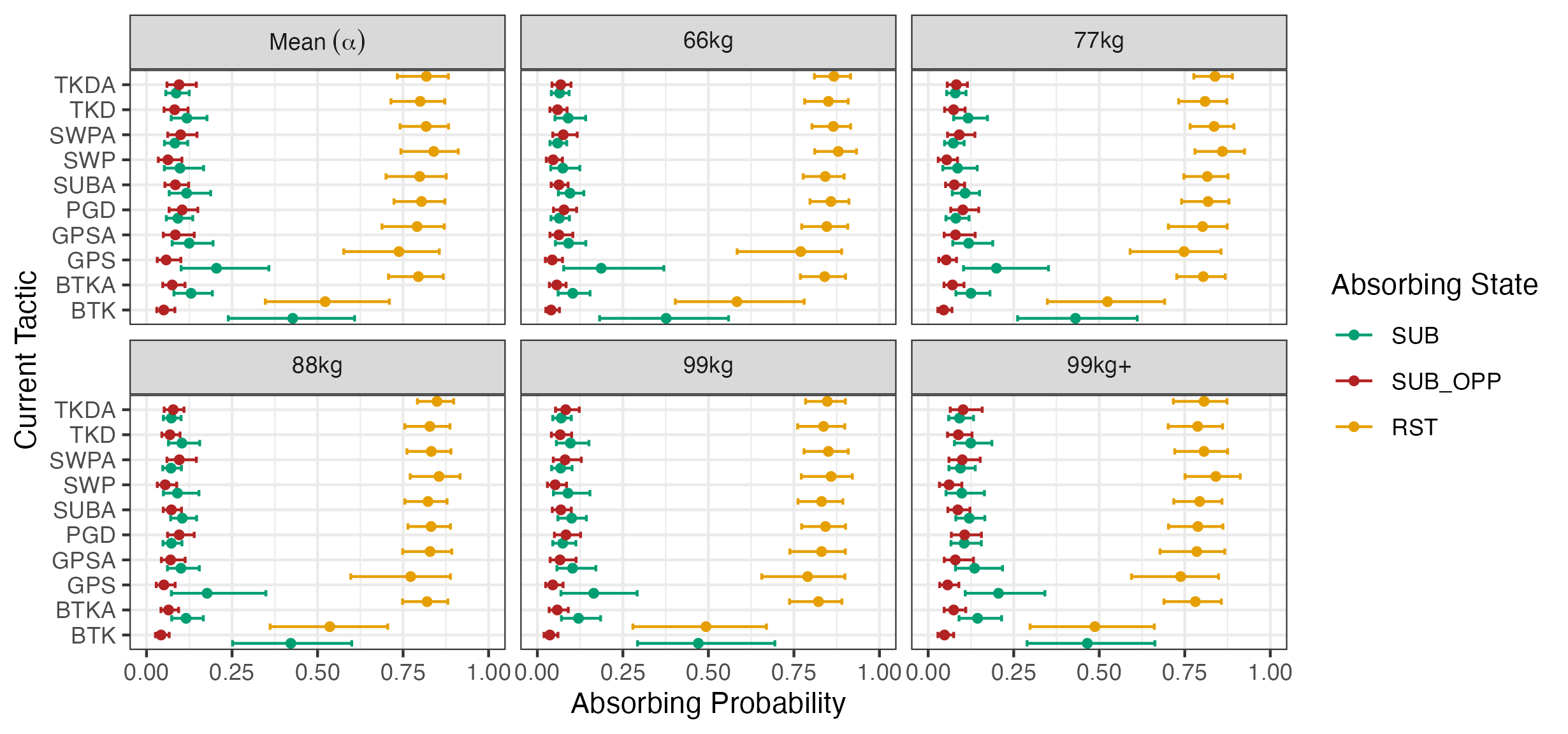}
    \caption{Posterior distributions of absorbing probabilities. The dots depict the posterior means which the error bars depict 95\% posterior credible intervals. Includes the overall mean estimate in the upper left, $\boldsymbol{\alpha}_i$, along with individual weight category distributions. Observe that an athlete is most likely to attain submission (SUB) when they have just performed a back-take (BTK): this is also the most variable tactical progression across weight categories.}
    \label{fig:prob_estimates}
\end{figure}

Fitting the EPA-HAMC model provides valuable insight. First we investigate the specific tactics of a given weight category by exploring the absorbing/eventual probabilities. All current tactics assume athlete ``1'' performed them, thus SUB represents that the same athlete attains submission while SUB\_OPP represents opponent obtaining a submission. Figure \ref{fig:prob_estimates} presents the posterior distribution of absorbing probabilities, offering an understanding of the relative efficiency of different tactics by weight category. Consistent with \cite{lamas2023no} we find that performing a BTK (back-take) is most likely to lead to attaining a submission. This tactic also shows the greatest variability across weight categories as seen in Figure~\ref{fig:prob_estimates}.

\begin{figure}
    \centering
    \includegraphics[width=1\linewidth]{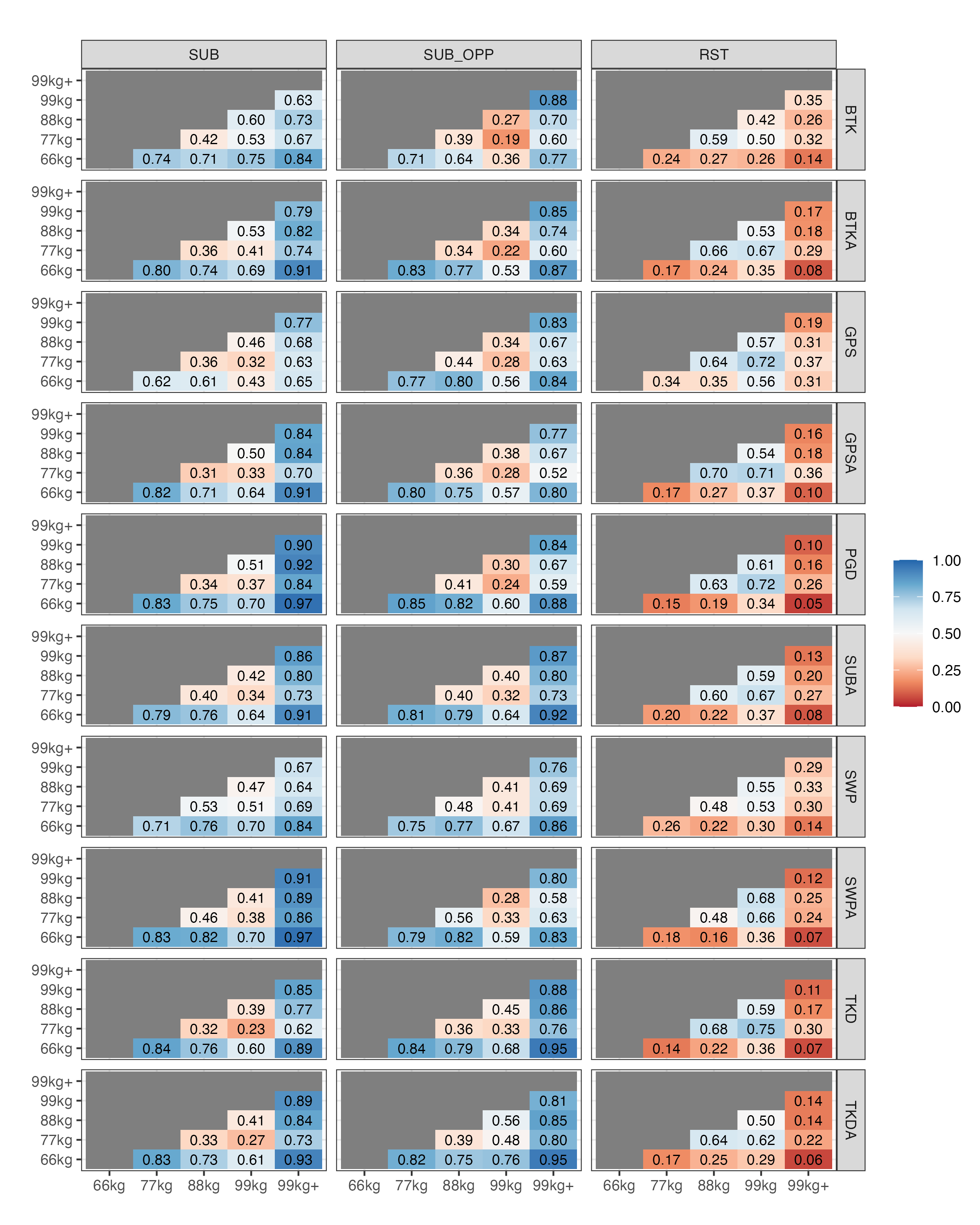}
    \caption{Posterior probability of pairwise comparisons that column weight category's absorbing probability is greater than row weight category absorbing probability. The darker reds and blues highlight larger posterior differences in absorbing probabilities. Weight categories 66kg and 99kg+ are consistently different from the other categories.}
    \label{fig:row_g_col}
\end{figure}

Observe further in Figure \ref{fig:prob_estimates} that there is a consistency in the pattern among all weight categories. This shows that although there are significant differences (Figure~\ref{fig:row_g_col}) present in the eventual probabilities, the underlying strategy of NGJJ is independent of weight category. We would expect this given that the same rules apply for each weight category.

We can sum the probabilities corresponding to each absorbing state within each weight category to give us an idea of the relative efficiency of each weight category. Efficiency represents how effective a given weight category is at attaining a given absorbing state overall. Table \ref{tab:efficiency} shows that 99kg+ is the most efficient in attaining submission. It also highlights that successfully performing a tactic is highly effective against counter-offense, as the opponent’s efficiency is notably low.

\begin{table}[h]
\centering
\footnotesize
\begin{tabular}{r|r|r|r|r|r}
  & 66kg & 77kg & 88kg & 99kg & 99kg+ \\ 
  \hline
SUB & 1.206 (0.727, 1.73) & 1.417 (0.962, 1.949) & 1.33 (0.89, 1.837) & 1.357 (0.89, 1.925) & 1.584 (1.071, 2.169) \\ 
  SUB\_OPP & 0.593 (0.384, 0.84) & 0.727 (0.491, 1.014) & 0.693 (0.473, 0.962) & 0.641 (0.415, 0.913) & 0.802 (0.54, 1.116) \\ 
  RST & 8.201 (7.468, 8.864) & 7.856 (7.097, 8.512) & 7.977 (7.239, 8.612) & 8.003 (7.221, 8.674) & 7.614 (6.764, 8.366) \\ 
\end{tabular}
\caption{Posterior means (and 95\% credible intervals) of relative efficiency of each weight category. Observe that 99kg+ is the most efficient at attaining a submission which the 66kg is the least efficient.}
\label{tab:efficiency}
\end{table}

Both Figure \ref{fig:prob_estimates} and Table \ref{tab:efficiency} show a reoccurring pattern in that the 66kg and 99kg+ differ the most from the other weight categories. For example, in Table \ref{tab:efficiency} the average efficiency of attaining a submission given you performed a successful tactic in 66kg is 1.206 while in 99kg+ it is 1.584 showing that in 99kg+ a submission is relatively more likely.

The hierarchical structure of the EPA-HAMC model allows us to examine the pairwise comparison of each weight category. Figure \ref{fig:row_g_col} illustrates how significantly different the weight categories are. The darker blue in the figure indicates that the probability of the given eventual transition is higher in the column weight category versus the row category, within a sub-matrix in Figure \ref{fig:row_g_col}. For example, when conditioning on TKD (take-down), we find that in weight category 99kg+, the focal competitor (Athlete ``1'') is more likely to eventually attain a submission than in the 66kg weight category, with posterior probability of 0.89. In NGJJ, tactics such as GPS and SWP are considered as middle tactics, meaning that they occur within the middle of a sequence and often do not directly lead to an decisive move, and their comparisons don't indicate as much of a significant difference between weight categories.

From Figure \ref{fig:row_g_col} we again observe the general trend that weight categories 66kg and 99kg+ are significantly different from the other categories, conditioning on all tactics. This trend suggests that a RST is much more probable in the 66kg compared to all others, and attaining a SUB is much more probable in the 99kg+.

Additionally we see that the model is able to overcome the prior on partitions in that when conditioned upon certain tactics such as SWPA, the 99kg category is much less likely to attain submission compared to 99kg+, despite being adjacent weight categories. We often see that weight categories 77kg, 88kg, and 99kg (three middle squares in each sub-matrix of Figure \ref{fig:row_g_col}) are not significantly different from one another in their eventual probability estimation.

\begin{figure}
    \centering
    \includegraphics[width=0.5\linewidth]{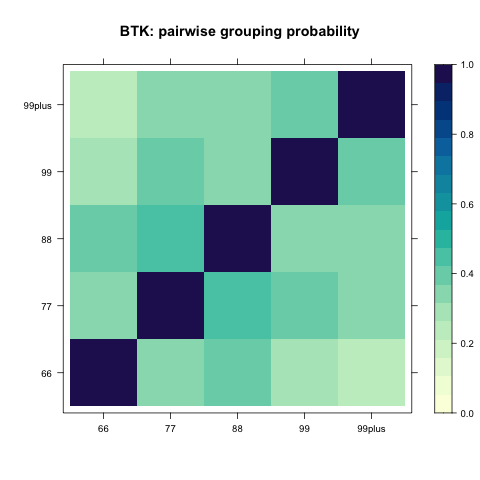}
    \caption{Posterior pairwise probabilities of co-clustering among weight groups for the transitions from back-take ($p_{BTK}^(k)$). The darker blue indicates that the given weight categories have a higher posterior pairwise probability of being clustered together. We observe that weight categories 77kg, 88kg and 99kg seem to be clustered together more often where 66kg and 99kg+ are more often on their own comparatively.}
    \label{fig:clustering_BTK}
\end{figure}

The random partition of the EPA-HAMC model allows us to observe pairwise similarities in immediate tactical progression between different weight categories. Figure \ref{fig:clustering_BTK} depicts the posterior probability of co-clustering among the five $\boldsymbol{p}_{BTK}^{(k)}$ vectors for transitions from BTK (back-take). For example, the probability vector from BTK in 66kg ($\boldsymbol{p}_{BTK}^{(66kg)}$) is clustered with the probability vector from BTK for be in 77kg ($\boldsymbol{p}_{BTK}^{(77kg)}$) in about 38\% of MCMC samples. Figure \ref{fig:clustering_BTK} can be interpreted as indicating how similar or dissimilar two respective weight categories are in their immediate progression from BTK. In NGJJ, performing BTK is considered a more influential tactic in leading to a submission. Figure \ref{fig:clustering_BTK} suggests some similarity between all weight categories in immediate progression, with 77kg, 88kg and 99kg being more similar while 66kg and 99kg+ are more often clustered on their own compared to others. Figure~\ref{fig:clustering_BTK} again illustrates the pattern that 66kg and 99kg+ are different in their tactical progression. When exploring the immediate tactical progression ($\boldsymbol{P}^{(k)}$) we do not notice any stark statistically significant differences among the weight categories of competition; this is illustrated in Section F of the Supplementary Material.

\section{Discussion}

We proposed a novel approach to analyzing sparse absorbing Markov chain data with inherent group structure. The HAMC method is also a novel approach, though it doesn't do as well when handling sparse data. However, it does allow for group specific estimation. EPA-HAMC demonstrates strong performance in posterior estimation with low cost of model misspecification. The model also introduces a new level of inference into these data, allowing for exploring the allocation of groupings given a current state as well as attaining group specific inference, together with a mean estimate.

Our analysis of NGJJ revealed that a back-take is most likely to lead to attaining a submission across all weight categories, consistent with findings from \cite{lamas2023no}. Further, we identified that the 66kg and 99kg+ weight categories differ significantly in their tactical progression to absorbing states compared to the other weight categories. 


Overall, the proposed methods provide the capacity to draw deeper inferences from absorbing Markov chain data than previously available. Future work will focus on exploring application to additional datasets and related domains.

\section*{Data and Code availability}
The code and NGJJ dataset underlying this article are publicly available on GitHub at \href{https://github.com/brighalvy/amcModels}{github.com/brighalvy/amcModels}. Instructions for reproduction and package dependencies are detailed in the repository's documentation.

\section*{Acknowledgment}
  Portions of this research were conducted with the advanced computing resources provided by Texas A\&M Department of Statistics Arseven Computing Cluster.

  The authors acknowledge the use of Google Gemini for proofreading and grammatical refinement of the manuscript, and OpenAI's Codex model for verifying programming code. All scientific code and final manuscript text were thoroughly checked and validated by the authors, who take full responsibility for the content of the publication.


\bibliography{references}

\appendix

\section{Example of structure of $\boldsymbol{P}$}
\label{sec:example_P}

Below is an example of the structure of $\boldsymbol{P}$ that utilizes six total transitional tactics with two absorbing states resulting in an 8x8 matrix of transition probabilities. We will let our transitional states have three unique tactics (A, B, C) that are performed by person 1 or 2, then we have an absorbing tactic (D). Let C1 mean that person 1 performs tactic C. Within this example we demonstrate the symmetry described within the NGJJ data, in that the assignment of person 1 or 2 is arbitrary. Thus in this example states 1-3 and states 4-6 are mirrors of each other. For example $p_{B1, A1}$ is the probability of person 1 performing tactic $A$ given they just performed tactic $B$ and $p_{B1, A1} = p_{B2, A2}$. Additionally $p_{C1, D2}$ is the probability of person 2 performing the absorbing tactic $D$ given that person 1 just performed tactic $C$ and $p_{C1, D2} = p_{C2, D1}$. The described symmetry is demonstrated through color within the matrix. Note that states seven and eight are the absorbing states and the structural ones and zeros can be seen in their corresponding rows.

\[
    \begin{blockarray}{r cccccccc}
        & \text{A1} & \text{B1} & \text{C1} & \text{A2} & \text{B2} & \text{C2} & \text{D1} & \text{D2} \\
        \begin{block}{r (cccccccc)}
            \text{A1} & \textcolor{blue}{p_{A1,A1}} & \textcolor{blue}{p_{A1,B1}} & \textcolor{blue}{p_{A1,C1}} & \textcolor{red}{p_{A1,A2}} & \textcolor{red}{p_{A1,B2}} & \textcolor{red}{p_{A1,C2}} & \textcolor{blue}{p_{A1,D1}} & \textcolor{red}{p_{A1,D2}} \\
            \text{B1} & \textcolor{blue}{p_{B1,A1}} & \textcolor{blue}{p_{B1,B1}} & \textcolor{blue}{p_{B1,C1}} & \textcolor{red}{p_{B1,A2}} & \textcolor{red}{p_{B1,B2}} & \textcolor{red}{p_{B1,C2}} & \textcolor{blue}{p_{B1,D1}} & \textcolor{red}{p_{B1,D2}} \\
            \text{C1} & \textcolor{blue}{p_{C1,A1}} & \textcolor{blue}{p_{C1,B1}} & \textcolor{blue}{p_{C1,C1}} & \textcolor{red}{p_{C1,A2}} & \textcolor{red}{p_{C1,B2}} & \textcolor{red}{p_{C1,C2}} & \textcolor{blue}{p_{C1,D1}} & \textcolor{red}{p_{C1,D2}} \\
            \text{A2} & \textcolor{red}{p_{A2,A1}} & \textcolor{red}{p_{A2,B1}} & \textcolor{red}{p_{A2,C1}} & \textcolor{blue}{p_{A2,A2}} & \textcolor{blue}{p_{A2,B2}} & \textcolor{blue}{p_{A2,C2}} & \textcolor{red}{p_{A2,D1}} & \textcolor{blue}{p_{A2,D2}} \\
            \text{B2} & \textcolor{red}{p_{B2,A1}} & \textcolor{red}{p_{B2,B1}} & \textcolor{red}{p_{B2,C1}} & \textcolor{blue}{p_{B2,A2}} & \textcolor{blue}{p_{B2,B2}} & \textcolor{blue}{p_{B2,C2}} & \textcolor{red}{p_{B2,D1}} & \textcolor{blue}{p_{B2,D2}} \\
            \text{C2} & \textcolor{red}{p_{C2,A1}} & \textcolor{red}{p_{C2,B1}} & \textcolor{red}{p_{C2,C1}} & \textcolor{blue}{p_{C2,A2}} & \textcolor{blue}{p_{C2,B2}} & \textcolor{blue}{p_{C2,C2}} & \textcolor{red}{p_{C2,D1}} & \textcolor{blue}{p_{C2,D2}} \\
            \text{D1} & 0 & 0 & 0 & 0 & 0 & 0 & 1 & 0 \\
            \text{D2} & 0 & 0 & 0 & 0 & 0 & 0 & 0 & 1 \\
        \end{block}
    \end{blockarray}
\]

\section{Full Joint Distribution}
\label{sec:full_joint}

We find the full joint distribution for the $i^{th}$ row (tactic) using the EPA-HAMC model in Equation 10:

\begin{align}
    &f(\{\boldsymbol{n}_i^{(k)}\}_{k=1}^{K}, \{\boldsymbol{\theta}_i^{(g)}\}_{g=1}^{G_i}, \gamma_i, \boldsymbol{\alpha}_i, \pi_i, \beta_i, \delta_i, \boldsymbol{\sigma}_i)  \nonumber\\
    &\quad =\left[ \prod_{g = 1}^{G_i}\prod_{j=1}^J\left(\theta_{i,j}^{(g)}\right)^{\sum_{k \in S_{g,i}}n_{i,j}^{(k)}} \right]\left[ \prod_{g = 1}^{G_i}f(\boldsymbol{\theta}_i^{(g)} \mid \gamma_i,\boldsymbol{\alpha}_i) \right]  f(\gamma_i)f(\boldsymbol{\alpha}_i) f(\pi_i \mid \beta_i, \delta_i, \boldsymbol{\sigma}_i)f(\beta_i)f(\delta_i)f(\boldsymbol{\sigma}_i)  \nonumber\\
    &=f(\gamma_i)f(\boldsymbol{\alpha}_i)f(\pi_i \mid \beta_i, \delta_i, \boldsymbol{\sigma}_i)f(\beta_i)f(\delta_i)f(\boldsymbol{\sigma}_i)\prod_{g = 1}^{G_i}\frac{1}{B(\gamma_i\boldsymbol{\alpha}_i)} \underbrace{\prod_{j=1}^J\left(\theta_{i,j}^{(g)}\right)^{\sum_{k \in S_{g,i}}n_{i,j}^{(k)} + \gamma_i\alpha_{i,j}}}_{\text{Dirichlet Kernel}},
    \label{eq:dirichlet_kernel}
\end{align}
\noindent with $B(\boldsymbol{a}) = \frac{\prod_{l=1}^{L}\Gamma(a_l)}{\Gamma(\sum_{l=1}^{L}a_l)}$, $\boldsymbol{a} = (a_1, a_2, ..., a_L)$. Also note for the application to NGJJ that $J = J_i$, because the number of possible transitions becomes row dependent. Letting,
\begin{align*}
    f(\gamma_i) &= \text{Gamma}(\gamma_i \mid \tau, \eta)\\
    f(\boldsymbol{\alpha_i}) &= \text{Dirichlet}(\boldsymbol{\alpha}_i \mid \frac{1}{J}\boldsymbol{1}_i) \\
    f(\pi_i \mid \beta_i, \delta_i, \boldsymbol{\sigma}_i) &= \text{EPA}(\pi_i \mid \beta_i, \delta_i, \lambda_i, \boldsymbol{\sigma}_i)\\
    f(\beta_i) &= \text{Gamma}(\beta_i \mid 3, 5)\\
    f(\delta_i) &= \omega \text{Beta}(\delta_i \mid .05, 20) + (1 - \omega)\text{Beta}(\delta_i \mid a, b)\\\
    f(\boldsymbol{\sigma}_i) &\propto 1
\end{align*}

\section{Marginal Conditional Distributions}
\label{sec:marg_cond}

We integrated out $\boldsymbol{\theta}_i^{(g)}$ to remove an unknown variable and obtain the joint distribution of the count data, and our remaining parameters. The integration process used uses the Dirichlet kernel (indicated in Equation \ref{eq:dirichlet_kernel}) for each $g \in \{1, ..., G_i\}$. It is found that each normalizing constant is $B(\boldsymbol{n}_{i}^{(k)} + \gamma_i\boldsymbol{\alpha}_{i})$, where $B(\cdot)$ is the Beta function. 

\vspace{-3mm}
\begin{align*}
    &f(\{\boldsymbol{n}_i^{(k)}\}_{k=1}^{K}, \gamma_i, \boldsymbol{\alpha}_i, \pi_i, \beta_i, \delta_i, \boldsymbol{\sigma}_i) \\
    &\quad =f(\gamma_i)f(\boldsymbol{\alpha}_i)f(\pi_i \mid \beta_i, \delta_i, \boldsymbol{\sigma}_i)f(\beta_i)f(\delta_i)f(\boldsymbol{\sigma}_i)\prod_{g = 1}^{G_i}\frac{1}{B(\gamma_i\boldsymbol{\alpha}_i)}B(\sum_{k \in S_{g,i}}\boldsymbol{n}_{i}^{(k)} + \gamma_i\boldsymbol{\alpha}_{i}) \\
    &\quad =f(\gamma_i)f(\boldsymbol{\alpha}_i)f(\pi_i \mid \beta_i, \delta_i, \boldsymbol{\sigma}_i)f(\beta_i)f(\delta_i)f(\boldsymbol{\sigma}_i)\prod_{g = 1}^{G_i}\frac{\Gamma(\gamma_i)}{\Gamma(\sum_{j=1}^{J}\sum_{k \in S_{ig}}n_{i,j}^{(k)}+\gamma_i)}\prod_{j=1}^{J}\frac{\Gamma(\sum_{k \in S_{ig}}n_{i,j}^{(k)}+\gamma_i\alpha_{i,j})}{\Gamma(\gamma_i\alpha_{i,j})}.
\end{align*}

\subsection{Marginal Conditional of $\pi_i$}

\begin{align*}
    &f(\pi_i \mid \{\boldsymbol{n}_i^{(k)}\}_{k=1}^{K}, \gamma_i, \boldsymbol{\alpha}_i, \beta_i, \delta_i, \boldsymbol{\sigma}_i) \\
    &\quad \propto f(\pi_i \mid \beta_i, \delta_i, \boldsymbol{\sigma}_i) \prod_{g = 1}^{G_i}\frac{\Gamma(\gamma_i)}{\Gamma(\sum_{j=1}^{J}\sum_{k \in S_{ig}}n_{i,j}^{(k)}+\gamma_i)}\prod_{j=1}^{J}\frac{\Gamma(\sum_{k \in S_{ig}}n_{i,j}^{(k)}+\gamma_i\alpha_{i,j})}{\Gamma(\gamma_i\alpha_{i,j})}
\end{align*}

\subsection{Marginal Conditional of $\boldsymbol{\sigma}_i$}

\begin{equation*}
    f(\boldsymbol{\sigma}_i \mid \cdot) \propto f(\pi_i \mid \beta_i, \delta_i, \boldsymbol{\sigma}_i)
\end{equation*}

\subsection{Marginal Conditional of $\beta_i$}

\begin{equation*}
    f(\beta_i \mid \cdot) \propto f(\pi_i \mid \beta_i, \delta_i, \boldsymbol{\sigma}_i)f(\beta_i)
\end{equation*}

\subsection{Marginal Conditional of $\delta_i$}

\begin{equation*}
    f(\delta_i \mid \cdot) \propto f(\pi_i \mid \beta_i, \delta_i, \boldsymbol{\sigma}_i)f(\delta_i)
\end{equation*}

\subsection{Marginal Conditional of $\gamma_i$}

\begin{equation*}
    f(\gamma_i \mid \{\boldsymbol{n}_i^{(k)}\}_{k=1}^{K}, \boldsymbol{\alpha}_i, \pi_i, \beta_i, \delta_i, \boldsymbol{\sigma}_i) \propto f(\gamma_i)\prod_{g = 1}^{G_i}\frac{\Gamma(\gamma_i)}{\Gamma(\sum_{j=1}^{J}\sum_{k \in S_{ig}}n_{i,j}^{(k)}+\gamma_i)}\prod_{j=1}^{J}\frac{\Gamma(\sum_{k \in S_{ig}}n_{i,j}^{(k)}+\gamma_i\alpha_{i,j})}{\Gamma(\gamma_i\alpha_{i,j})}
\end{equation*}

\subsection{Marginal Conditional of $\alpha_i$}

\begin{equation*}
    f(\boldsymbol{\alpha}_i \mid \{\boldsymbol{n}_i^{(k)}\}_{k=1}^{K}, \gamma_i, \pi_i, \beta_i, \delta_i, \boldsymbol{\sigma}_i) \propto f(\boldsymbol{\alpha}_i)\prod_{g = 1}^{G_i}\frac{\Gamma(\gamma_i)}{\Gamma(\sum_{j=1}^{J}\sum_{k \in S_{ig}}n_{i,j}^{(k)}+\gamma_i)}\prod_{j=1}^{J}\frac{\Gamma(\sum_{k \in S_{ig}}n_{i,j}^{(k)}+\gamma_i\alpha_{i,j})}{\Gamma(\gamma_i\alpha_{i,j})}
\end{equation*}

\subsection{Full Conditional of $\boldsymbol{p}_i^{(k)}$}

Recall that $\boldsymbol{p}_i^{(k)} = \sum_{g = 1}^G \boldsymbol{\theta}_i^{(g)}\cdot \boldsymbol{I}\{k \in S_{g,i}\}$. Thus we just need to find the full conditional of $\boldsymbol{\theta}^{(g)}$. Give $\pi_i$ we use the Markov chain - Dirichlet conjugacy to find:

\begin{equation*}
    f(\boldsymbol{\theta}_i^{(g)} \mid \{\boldsymbol{n}_i^{(k)}\}_{k=1}^{K}, \gamma_i, \boldsymbol{\alpha}_i, \pi_i, \beta_i, \delta_i, \boldsymbol{\sigma}_i) = \text{Dirichlet}(\boldsymbol{\theta}_i^{(g)} \mid \gamma_i\boldsymbol{\alpha}_i + \sum_{k \in S_{g,i}}\boldsymbol{n}_i^{(k)}).
\end{equation*}

\section{Markov Chain - Monte Carlo (MCMC)}
\label{sec:app_MCMC}

The following steps are taken in our MCMC algorithm:

\begin{enumerate}[label=\textbf{\underline{Step \arabic*:}}]

\item Draw $\pi_i$:

For each group we find the marginal conditional probability of that group joining each of the existing clusters of the remaining groups or its own new group as in \cite{dahl2017random}.

\item Draw $\boldsymbol{\sigma}_i$:

This is done using Metropolis-Hastings algorithm with a uniform proposal scheme outlined by \cite{dahl2017random}.

\item Draw $\beta_i$:

Using a classic Metropolis-Hastings algorithm with a Gaussian proposal scheme.

\item Draw $\delta_i$:

Using a classic Metropolis-Hastings algorithm with a Gaussian proposal scheme.

\item Draw $\gamma_i$:

Let \[\mathcal{L}(\gamma) = \prod_{g = 1}^{G_i}\frac{\Gamma(\gamma)}{\Gamma(\sum_{j=1}^{J}\sum_{k \in S_{ig}}n_{i,j}^{(k)}+\gamma)}\prod_{j=1}^{J}\frac{\Gamma(\sum_{k \in S_{ig}}n_{i,j}^{(k)}+\gamma\alpha_{i,j})}{\Gamma(\gamma\alpha_{i,j})}\] be the marginal likelihood evaluated at $\gamma$ and,
\begin{equation*}
    g(\gamma) = f(\gamma)\mathcal{L}(\gamma)
\end{equation*} where $f()$ is the prior on $\gamma_i$. The following quantile slice sampler algorithm will be used for $\gamma_i$ \citep{heiner2023}.

\begin{algorithm}[H] 
\caption{$\gamma$ Update}
\begin{algorithmic}
\State $y \gets \text{runif}(1,0,g(\gamma)/f(\gamma))$
\State $L \gets 0$
\State $R \gets 1$
\State $u \gets \text{runif}(1, L, R)$
\State $\gamma_{new} \gets F^{-1}(u)$
\While{$y > g(\gamma_{new})/f(\gamma_{new})$}
    \If{$\gamma_{new} < \gamma$}
        \State $L \gets u$
    \Else 
        \State $R \gets u$
    \EndIf
    \State $u \gets \text{runif}(1, L, R)$
    \State $\gamma_{new} \gets F^{-1}(u)$
\EndWhile
\end{algorithmic}
\end{algorithm}
\noindent We use $F^{-1}(x)$ to denote the quantile function of the Gamma prior evaluated at $x$.

\item Draw $\boldsymbol{\alpha}_i$:

We use the method outline in Section 2.4 of the main manuscript.

\item Draw $\boldsymbol{p}_i^{(k)}$:

We do this by drawing the corresponding $\boldsymbol{\theta}_i^{(g)}$ from the $\text{Dirichlet}(\gamma_i\boldsymbol{\alpha}_i + \sum_{k \in S_{g,i}}\boldsymbol{n}_i^{(k)})$.
\end{enumerate}

\section{Simulation 3: Exploring the Effect of $I$ and $K$}
\label{sec:sim3}

\textbf{Simulation 3} explores the effect of varying the number of transitory states and groups on the reconstruction of $\boldsymbol{P}^{(k)}$. Data were simulated under \textit{Scenario 3} for each combination of $I \in \{5, 15, 30\}$ and $K \in \{5, 15\}$ to observe the effect of $I$ and $K$ on recovering $\boldsymbol{P}^{(k)}$ within \textit{Scenario 3}. We differ from the data generating strategy given in Section 3 of the main manuscript in that we no longer use each transitory state 10 times. Rather we will use values of $\{10, 5, 3\}$ as $I$ increases. The change in the number of times state is used to initialize a sequence allows the total sample size to remain proportional to $I$ and $K$. 

\begin{figure}[H]
    \centering
    \includegraphics[width=\linewidth]{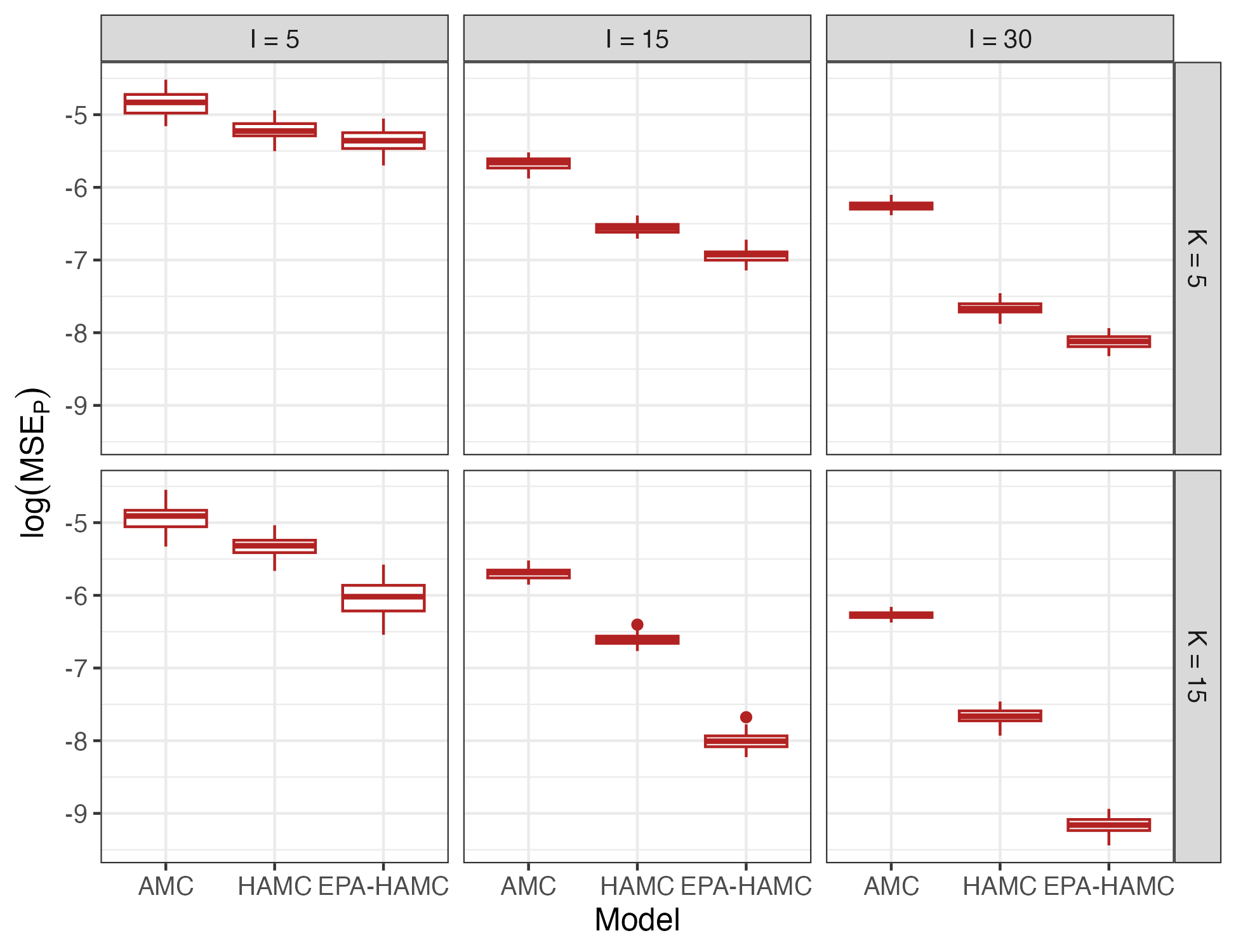}
    \caption{log(MSE$_P$) across all models for varying numbers of transitory states ($I$) and groups ($K$). These figures demonstrate that as both $I$ and $K$ increase, the models get more accurate in recovering the short term behavior.}
    \label{fig:MSE_P_sim2}
\end{figure}

\begin{figure}[H]
    \centering
    \includegraphics[width=\linewidth]{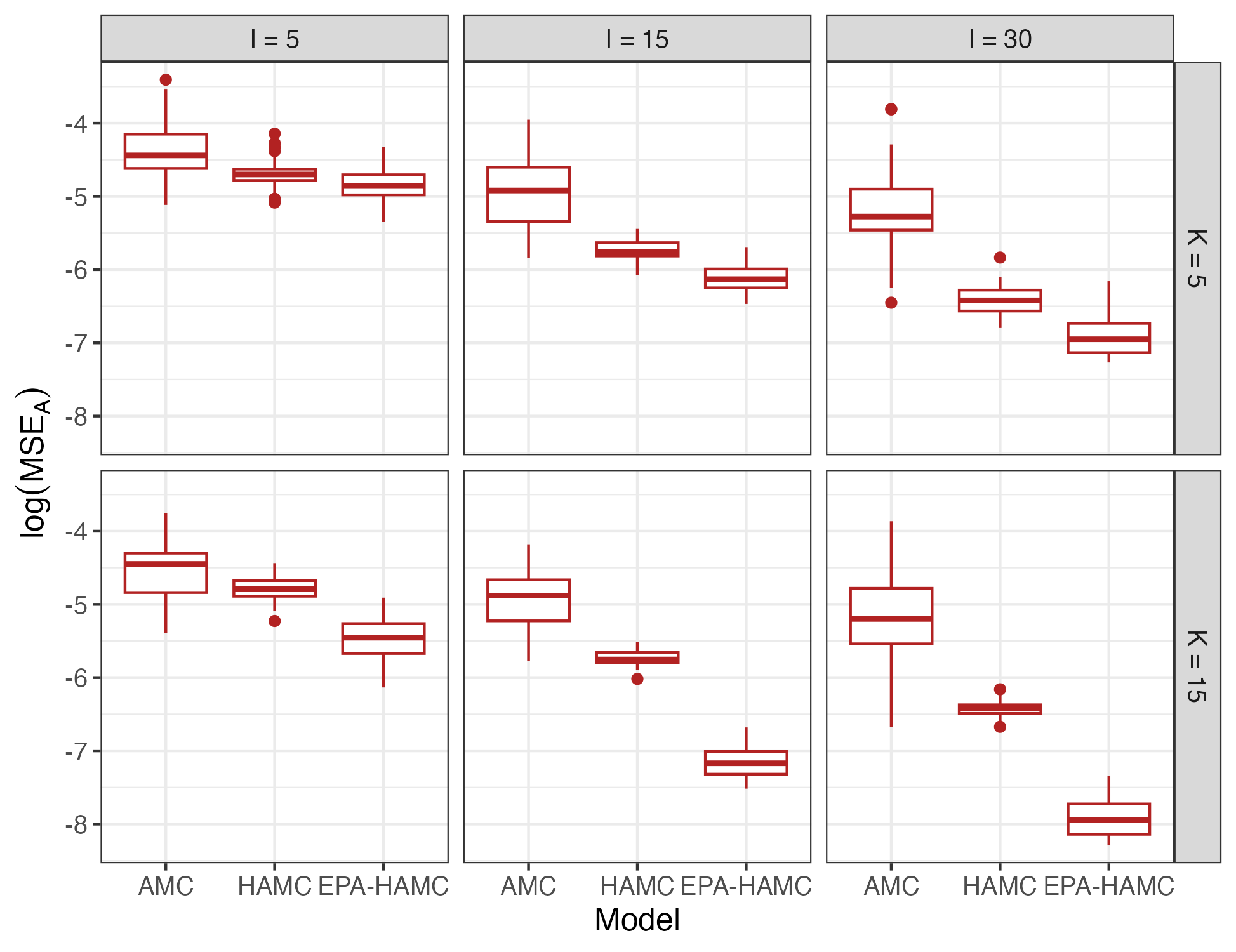}
    \caption{log(MSE$_A$) across all models for varying numbers of transitory states ($I$) and groups ($K$). These figures demonstrate that as both $I$ and $K$ increase the models gets more accurate in recovering the long-term behavior.}
    \label{fig:MSE_A_sim2}
\end{figure}

The results displayed in Figures \ref{fig:MSE_P_sim2} and \ref{fig:MSE_A_sim2} depict the log(MSE$_P$) and log(MSE$_A$). They show that as the number of transitory states ($I$) increases the overall estimation improves. The reduction in the MSE is expected. Firstly because increasing the number of states naturally decreases the average magnitude of individual transition probabilities, inherently bounding the magnitude of the squared errors. Second, as the sample size increases proportionally with $I$ we gain additional observations to help improve upon MSE.

Conversely, increasing the number of groups ($K$) results in minimal to no change in MSE for both immediate and long-term behavior. Because the simulation scales total sample size proportionally with $I$ and $K$, increasing $K$ maintains a constant sample size per group rather than increasing the information density per parameter. As a result we observe no change in the performance as $K$ increases.

\section{NGJJ Results}
\label{sec:app_ngjj}

\begin{figure}
\centering

\subcaptionbox{Fig.\ A\label{fig:Ex_Im}}{\includegraphics[width=.45\textwidth]{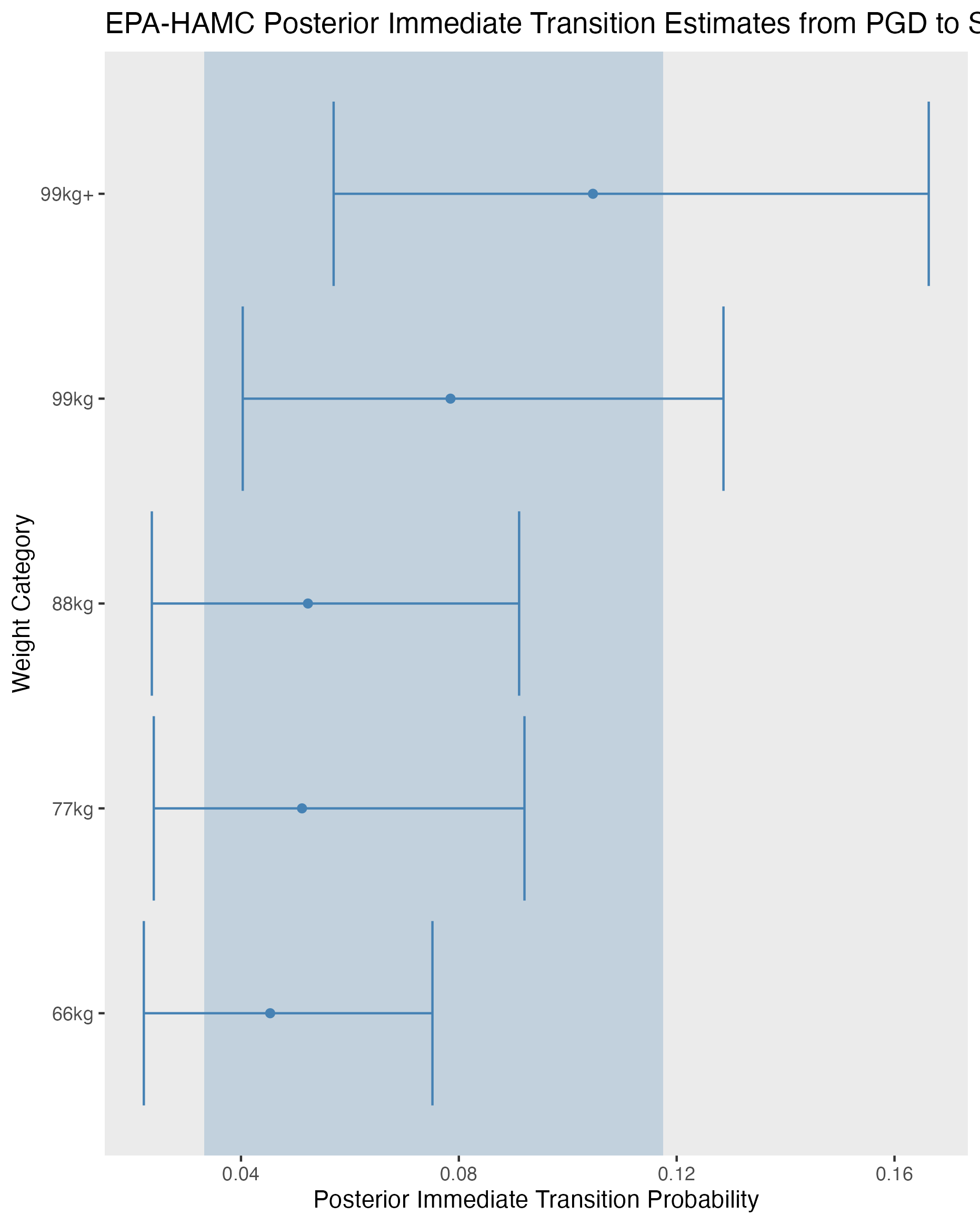}}\quad
\subcaptionbox{Fig.\ B\label{fig:Ex_Im2}}{\includegraphics[width=.45\textwidth]{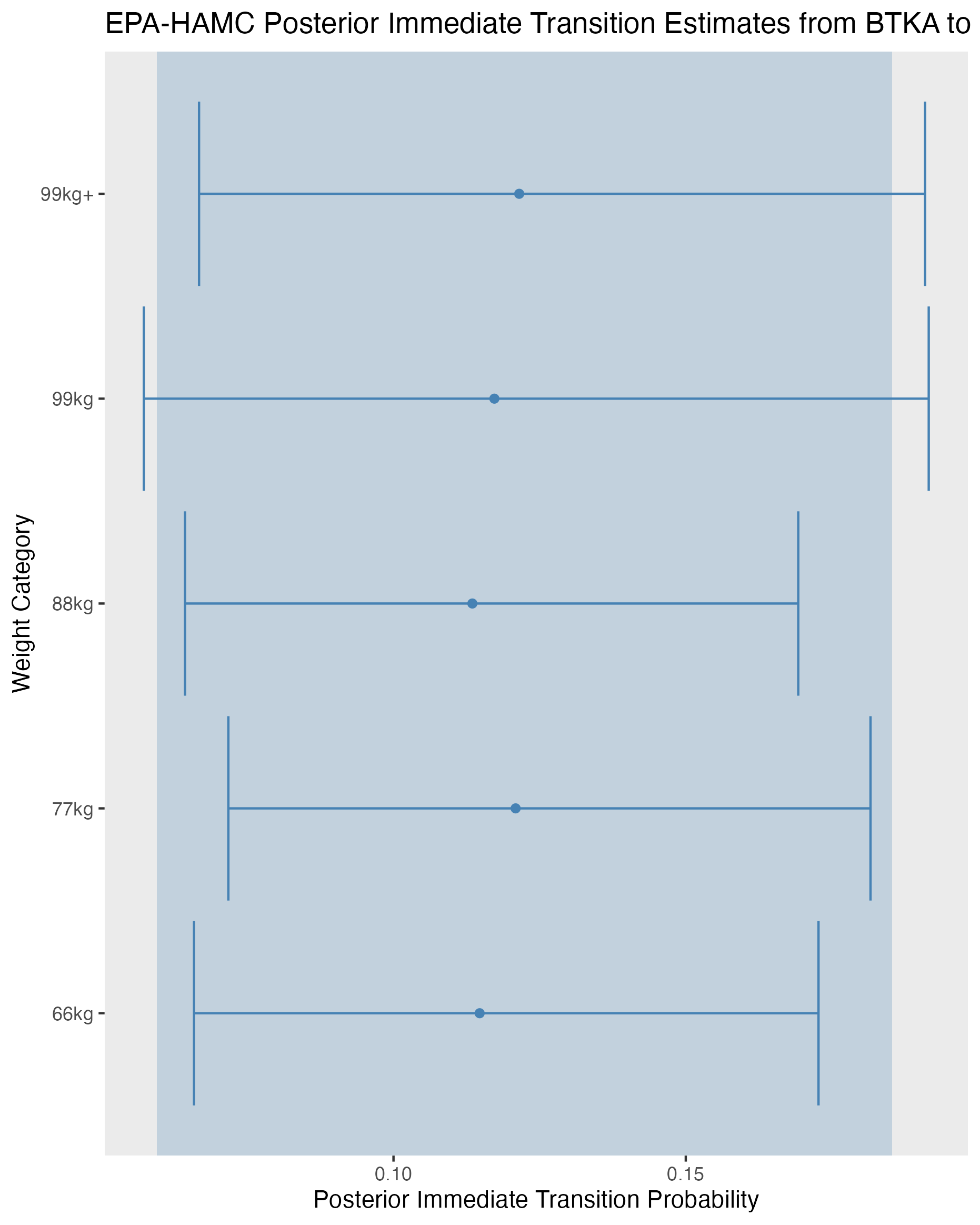}}
\subcaptionbox{Fig.\ C\label{fig:Ex_Im3}}{\includegraphics[width=.45\textwidth]{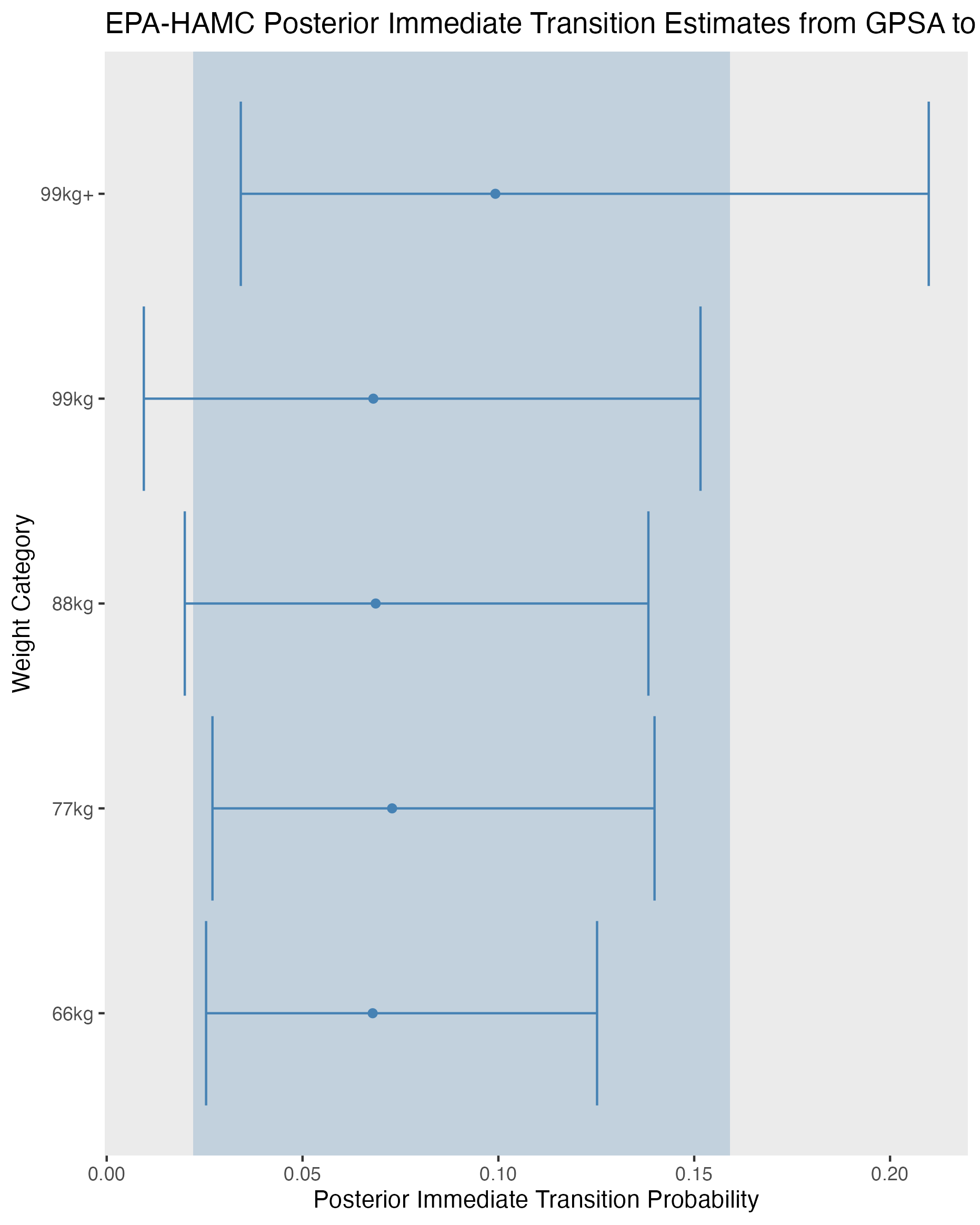}}
\subcaptionbox{Fig.\ D\label{fig:Ex_Im4}}{\includegraphics[width=.45\textwidth]{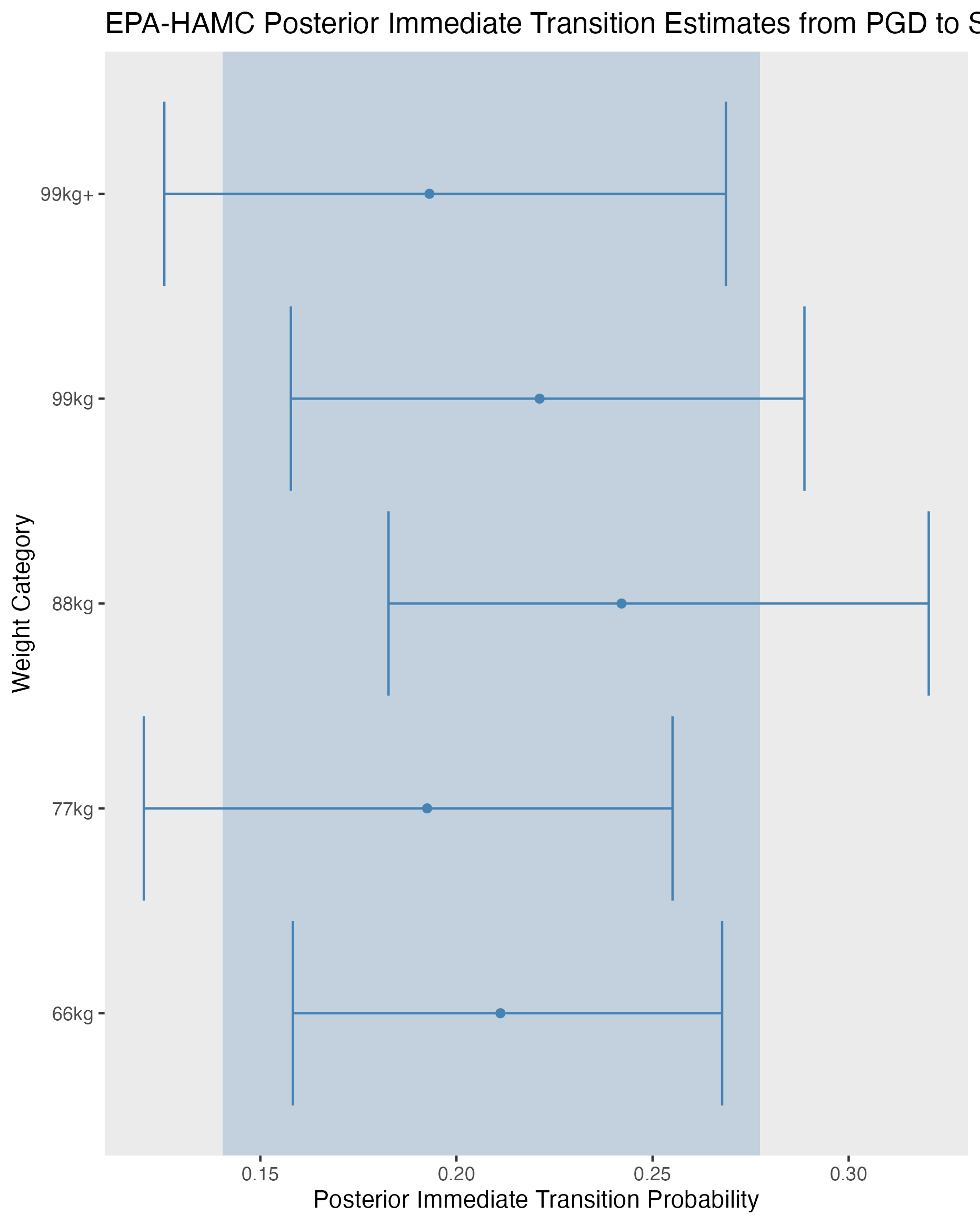}}
\subcaptionbox{Fig.\ E\label{fig:Ex_Im5}}{\includegraphics[width=.45\textwidth]{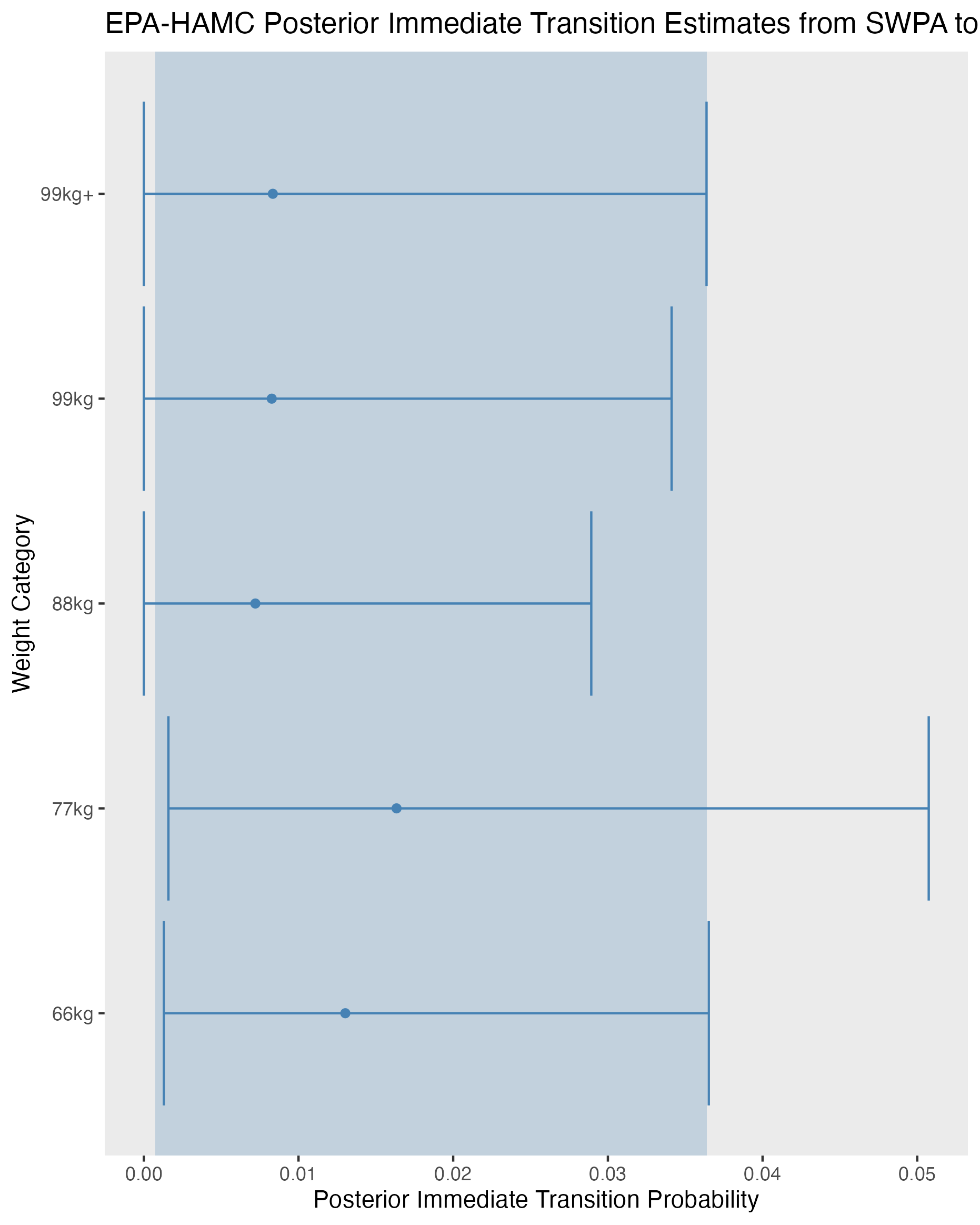}}
\subcaptionbox{Fig.\ F\label{fig:Ex_Im6}}{\includegraphics[width=.45\textwidth]{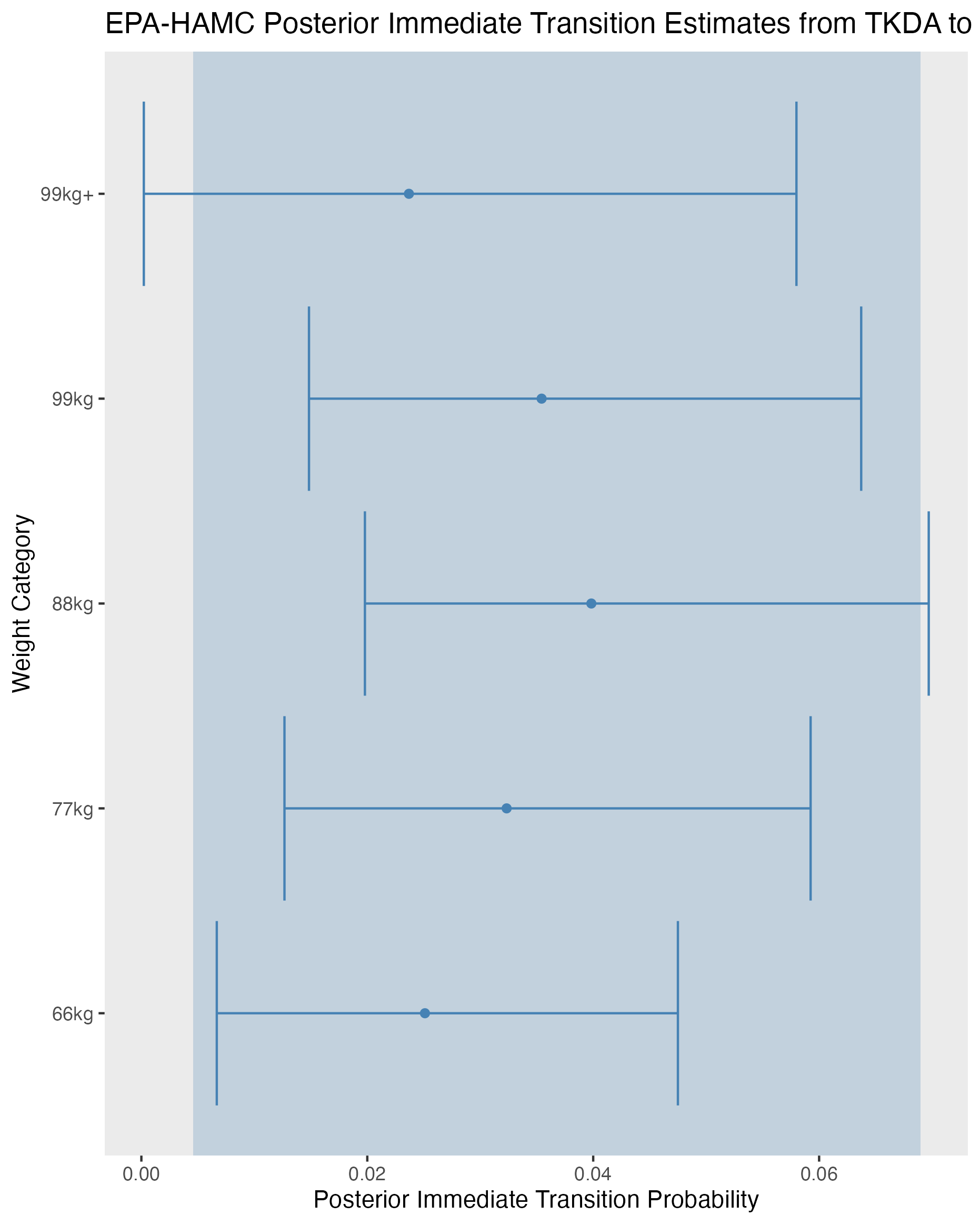}}

\caption{Posterior estimates of immediate transition probabilities with 95\% posterior credible intervals. The shaded regions on each plot are the mean 95\% credible intervals coming from estimates of $\boldsymbol{\alpha}_i$}
\label{fig:immediate_trans}
\end{figure}

\end{document}